\documentclass[epj,nopacs,twocolumn]{svjour}
\usepackage{graphicx}
\usepackage{amssymb}
\usepackage{amsmath}

\usepackage{comment}

\usepackage{subfigure}
\usepackage{setspace}
\usepackage{hyperref}
\usepackage{slashed}
\usepackage{bm}
\usepackage{afterpage}

\newcommand{\RWZPrecisionStatSyst}{1.4\%}
\newcommand{\RWZPrecisionSyst}{1.3\%}

\newcommand{\ttV}{\mbox{${t}\bar{t}V$}}

\newcommand{\RW}{\mbox{$R(W)$}}
\newcommand{\RbbWW}{\mbox{$R(bbWW)$}}
\newcommand{\RZ}{\mbox{$R(Z)$}}
\newcommand{\RWZ}{\mbox{$R(WZ)$}}

\newcommand{\RD}{\mbox{$R(D)$}}
\newcommand{\RDstar}{\mbox{$R(D^*)$}}

\newcommand{\ee}{\mbox{$e^+e^-$}}
\newcommand{\mumu}{\mbox{$\mu^+\mu^-$}}
\newcommand{\tautau}{\mbox{$\tau^+\tau^-$}}

\newcommand{\ellell}  {\mbox{$\ell^+ \ell^-$}}

\newcommand{\qq}  {\mbox{${q}\bar{q}$}}

\renewcommand{\tt}  {\mbox{${t}{t}$}}

\newcommand{\WW}  {\mbox{${W}^+{W}^-$}}

\newcommand{\Zee}{\mbox{$Z/\gamma^*$} \mbox{$\rightarrow$} \mbox{$ee$}}
\newcommand{\Zmumu}{\mbox{$Z/\gamma^*$} \mbox{$\rightarrow$} \mbox{$\mu\mu$}}
\newcommand{\Ztautau}{\mbox{$Z/\gamma^*$} \mbox{$\rightarrow$} \mbox{$\tau\tau$}}

\newcommand{\Zemu}  {\mbox{$Z$} \mbox{$\rightarrow$} \mbox{$\tau\tau$} \mbox{$\rightarrow$} \mbox{$ e\mu$}}
\newcommand{\Zetau}  {\mbox{$Z$} \mbox{$\rightarrow$} \mbox{$\tau\tau$} \mbox{$\rightarrow$} \mbox{$ e\tau_{\mathrm{\,had}}$}}
\newcommand{\Zmutau}  {\mbox{$Z$} \mbox{$\rightarrow$} \mbox{$\tau\tau$} \mbox{$\rightarrow$} \mbox{$ \mu\tau_{\mathrm{\,had}}$}}
\newcommand{\Zltau}  {\mbox{$Z$} \mbox{$\rightarrow$} \mbox{$\tau\tau$} \mbox{$\rightarrow$} \mbox{$ \ell\tau_{\mathrm{\,had}}$}}

\newcommand{\Zemutau}  {\mbox{$Z$} \mbox{$\rightarrow$} \mbox{$ ee, \mu\mu, \tau\tau$}}
\newcommand{\Ztaulept}  {\mbox{$Z$} \mbox{$\rightarrow$} \mbox{$\tau\tau$} \mbox{$\rightarrow$} \mbox{$\ell\tau_{\mathrm{\,lept}}$}}

\newcommand{\emu}  {\mbox{$e\mu$}}
\newcommand{\etau}  {\mbox{$e\tau_{\mathrm{\,had}}$}}
\newcommand{\mutau}  {\mbox{$\mu\tau_{\mathrm{\,had}}$}}
\newcommand{\ltau}  {\mbox{$\ell\tau_{\mathrm{\,had}}$}}

\newcommand{\ttWW}  {\mbox{${t}\bar{t}$} \mbox{$\rightarrow$} \mbox{$ b\bar{b}W^+W^-$}}
\newcommand{\ttWWltau}  {\mbox{$tt$} \mbox{$\rightarrow$} \mbox{$ bb\ell\tau_{\mathrm{\,had}}$}}

\newcommand{\ttWWemu}  {\mbox{$tt$} \mbox{$\rightarrow$} \mbox{$ bbe\mu$}}
\newcommand{\tWb}  {\mbox{$tWb$}}
\newcommand{\ttWWltaulept}  {\mbox{$tt$} \mbox{$\rightarrow$} \mbox{$ bb\ell\tau_{\mathrm{\,lept}}$}}

\newcommand{\ttotaunu}  {\mbox{$t$} \mbox{$\rightarrow$} \mbox{$ b\tau\nu$}}
\newcommand{\ttolnu}  {\mbox{$t$} \mbox{$\rightarrow$} \mbox{$ b\ell\nu$}}

\newcommand{\Wlnu}  {\mbox{$W$} \mbox{$\rightarrow$} \mbox{$\ell\nu$}}
\newcommand{\Wenu}  {\mbox{$W$} \mbox{$\rightarrow$} \mbox{$ e\nu$}}
\newcommand{\Wmunu}  {\mbox{$W$} \mbox{$\rightarrow$} \mbox{$\mu\nu$}}
\newcommand{\Wtaunu}  {\mbox{$W$} \mbox{$\rightarrow$} \mbox{$\tau\nu$}}
\newcommand{\Wtauhad}  {\mbox{$W$} \mbox{$\rightarrow$} \mbox{$\tau_{\mathrm{\,had}}\nu$}}

\newcommand{\Wtaunul}  {\mbox{$W$} \mbox{$\rightarrow$} \mbox{$\tau\nu$} \mbox{$\rightarrow$} \mbox{$\ell\nu\nu$}}

\newcommand{\WWlnutaunu}  {\mbox{$WW$} \mbox{$\rightarrow$} \mbox{$\ell\nu\tau_{\mathrm{\,had}}\nu$}}
\newcommand{\ZWlltaunu}  {\mbox{$ZW$} \mbox{$\rightarrow$} \mbox{$\ell\ell\tau_{\mathrm{\,had}}\nu$}}

\newcommand{\tauhad}   {\mbox{$\tau_{\mathrm{\,had}}$}}
\newcommand{\taulept}   {\mbox{$\tau_{\mathrm{\,lept}}$}}

\newcommand{\BtoTauD} {\mbox{$\overline{B}^0$} \mbox{$\rightarrow$} \mbox{$\tau^-\overline{\nu}_\tau D^+$}}
\newcommand{\BtoTauDstar} {\mbox{$\overline{B}^0$} \mbox{$\rightarrow$} \mbox{$\tau^-\overline{\nu}_\tau D^{*+}$}}
\newcommand{\BtoTauDandstar} {\mbox{$\overline{B}^0$} \mbox{$\rightarrow$} \mbox{$\tau^-\overline{\nu}_\tau D^{(*)+}$}}
\newcommand{\BtoLD} {\mbox{$\overline{B}^0$} \mbox{$\rightarrow$} \mbox{$\ell^-\overline{\nu}_\ell D^+$}}
\newcommand{\BtoLDstar} {\mbox{$\overline{B}^0$} \mbox{$\rightarrow$} \mbox{$\ell^-\overline{\nu}_\ell D^{*+}$}}
\newcommand{\BtomuD} {\mbox{$\overline{B}^0$} \mbox{$\rightarrow$} \mbox{$\mu^-\overline{\nu}_\tau D^+$}}
\newcommand{\BtomuDstar} {\mbox{$\overline{B}^0$} \mbox{$\rightarrow$} \mbox{$\mu^-\overline{\nu}_\tau D^{*+}$}}
\newcommand{\BtoKee} {\mbox{$B^+$} \mbox{$\rightarrow$} \mbox{$ K^+e^+e^-$}}
\newcommand{\BtoKmumu} {\mbox{$B^+$} \mbox{$\rightarrow$} \mbox{$ K^+\mu^+\mu^-$}}
\newcommand{\BtoKstarmumu} {\mbox{$B^0$} \mbox{$\rightarrow$} \mbox{$ K^{*0}\mu^+\mu^-$}}

\newcommand{\tauenunu}  {\mbox{$\tau^-$} \mbox{$\rightarrow$} \mbox{$ e^-\bar{\nu}_e\nu_\tau$}}
\newcommand{\taumununu}  {\mbox{$\tau^-$} \mbox{$\rightarrow$} \mbox{$\mu^-\bar{\nu}_\mu\nu_\tau$}}

\newcommand{\phistar}{\mbox{$\phi^{*}_\eta$}}
\newcommand{\sts}{\mbox{$\sin\theta^*_\eta$}}

\newcommand{\sumcos}{\mbox{$\Sigma\cos\Delta\phi$}}

\newcommand{\pt}{\mbox{$p_T$}} 
\newcommand{\Zpt}{\mbox{$p_T(Z)$}} 
\newcommand{\mtt}{\mbox{$m(t\bar{t})$}}

\newcommand{\at}{\mbox{$a_T$}}
\newcommand{\al}{\mbox{$a_L$}}

\newcommand{\modeta}{\mbox{$|\eta|$}}

\newcommand{\mstar}{\mbox{$m^*_3$}}
\newcommand{\mtaujet}{\mbox{$m(j$$-$$\tauhad)$}}

\newcommand{\etmiss}{\mbox{$\slashed{E}_T$}}

\newcommand{\invfb}  {\mbox{fb$^{-1}$}}
\newcommand{\intL}{\mbox{$\int\!\! L\, \d t$}}

\renewcommand{\d}  {\mbox{\,d}}

\begin{document}

\title{A self-calibrating, double-ratio method to test tau lepton universality in W boson decays at the LHC}
\author{S. Dysch \and T. R. Wyatt.}
\institute{Particle Physics Group, Department of Physics and Astronomy, University of Manchester, UK.}
\titlerunning{A self-calibrating, double-ratio method to test tau lepton universality in W boson decays}
\authorrunning{S. Dysch and T. R. Wyatt.}
\date{Received: 26/10/2019}
%
\abstract{
  Measurements in $W^+W^-$ events at LEP2 and in $B$ hadron semileptonic decays at B factories and LHCb provide
 intriguing hints of a violation of lepton universality in the charged current coupling of
 tau leptons relative to those for electrons and muons.
 We propose a novel,  self-calibrating method to test tau lepton universality in $W$ boson decays at the LHC.
 We compare directly the ratio of the numbers of selected \ltau\ and \emu\ final states in di-leptonic top quark pair events with that in \Ztautau\ events. 
 Here $\ell=e\ \mathrm{or}\ \mu$ and \tauhad\ is a candidate semi-hadronic tau decay.
This ``double-ratio'' cancels to first order sensitivity to systematic uncertainties on the reconstruction of
$e$, $\mu$, and $\tau$ leptons, thus improving very significantly  the precision to which tau lepton universality can be tested in $W$
boson decay branching ratios at
the LHC.
Using particle-level Monte Carlo events, and a parameterised simulation of detector performance, we demonstrate the effectiveness of
the method and estimate  the most significant residual sources of uncertainty arising from experimental and phenomenological systematics.
Our studies indicate that a single LHC experiment precision on the tau lepton universality test of around \RWZPrecisionStatSyst\ is achievable with a data set of \intL~=~140~\invfb\ at
$\sqrt{s}$~=~13~TeV.
This would improve significantly upon the precision of 2.5\% on the four-experiment combined LEP2 measurements.
If the central value of the proposed new measurement were
equal to the central value of the LEP2 measurement this would yield an observation of BSM physics at a significance level
of around 5$\sigma$.
} 

%
\maketitle
%

\section{Introduction}    
\label{sec:intro}

An important feature of the {\it Standard Model} (SM) of particle physics is {\it Lepton Universality} (LU): the idea that 
the  {\it Electroweak} (EW)  couplings of the leptons are identical in each
fermion generation.
For the {\it Neutral Current} (NC) interactions, mediated in the SM by the $\gamma$ and $Z$ bosons, the validity of LU for the
three flavours of charged leptons ($e, \mu, \tau)$ has been
demonstrated at the scale of the $Z$ boson mass $m_Z$ at LEP1 and SLC to high precision~\cite{bib:LEP1}.
For example, the ratios of the leptonic partial widths ($\Gamma$) or branching fractions ($\mathcal{B}$) of the $Z$ boson are:
\begin{equation}
\begin{array}{lllll}
 \displaystyle  \frac{\Gamma_{\mu\mu}}{\Gamma_{ee}} &\equiv & \displaystyle \frac{\mathcal{B}(\Zmumu)}{\mathcal{B}(\Zee)} &=& 1.0009 \pm 0.0028, \\ \\
 \displaystyle  \frac{\Gamma_{\tau\tau}}{\Gamma_{ee}} &\equiv & \displaystyle \frac{\mathcal{B}(\Ztautau)}{\mathcal{B}(\Zee)} &=& 1.0019 \pm 0.0032,
\end{array}
\label{eqn:Z}
\end{equation}
\noindent which are consistent with LU to a precision of around three per mille~\cite{bib:LEP1}.
Measurements of the leptonic asymmetry parameters $\mathcal{A}_\ell,$ from forward-backward asymmetries, the left-right asymmetry,
and the tau polarisation and its asymmetry:
\begin{equation}
\begin{array}{lll}
 \displaystyle  \mathcal{A}_e &=& 0.1514 \pm 0.0019, \\ \\
 \displaystyle  \mathcal{A}_\mu &=& 0.1456 \pm 0.0091, \\ \\
 \displaystyle  \mathcal{A}_\tau &=& 0.1449 \pm 0.0040, \\
\end{array}
\end{equation}
\noindent are also consistent with LU, albeit at the precision of a few percent~\cite{bib:LEP1}.

In contrast, for the  {\it Charged Current} (CC) interactions, mediated in the SM by the $W^\pm$ bosons, the experimental
measurements of leptonic branching ratios are less precise.
A direct test of LU at the scale of the  $W$ boson mass $m_W$ at LEP2 can be made from the ratios of the leptonic branching
fractions of the $W$ boson.
The ratio~\cite{bib:LEP2}:
\begin{equation}
 \frac{\mathcal{B}(\Wmunu)}{\mathcal{B}(\Wenu)} = 0.993 \pm 0.019
\end{equation}
is consistent with $e$-$\mu$ universality to a precision  of around two percent.
However a hint at the possible violation of LU in the CC couplings of the $\tau$ is present in the ratio~\cite{bib:LEP2}:
\begin{equation}
 \RW \equiv \frac{\mathcal{B}(\Wtaunu)}{\mathcal{B}(\Wlnu)} = 1.066 \pm 0.025,
\end{equation}
where $\mathcal{B}(\Wlnu)$ is the average of the branching fractions for $\Wenu$ and $\Wmunu$.
This result deviates from the assumption of $\tau$-$\ell$
universality\footnote{In this paper the symbol $\ell$ is taken to denote an electron ($e$) or a muon ($\mu$), but not a tau ($\tau$)
  lepton.} by 2.6$\sigma$ (standard deviations).

At lower mass scales $e$-$\mu$ universality is tested very precisely, for example in leptonic $\tau$ decays the ratio of the muonic
and electronic partial widths is measured to be~\cite{bib:pdg}: 
\begin{equation}
  \frac{\Gamma(\taumununu)}{\Gamma(\tauenunu)} = 0.9762 \pm 0.0028,
\end{equation}
which is consistent with the SM  prediction including mass effects of 0.9726~\cite{bib:pdg}.
$e$-$\mu$ universality is also tested in the decays of charged kaons~\cite{bib:pdg}:
\begin{equation}
	\frac{\Gamma(K^-\to e^-\bar{\nu}_e)}{\Gamma(K^-\to\mu^-\bar{\nu}_\mu)}=(2.488\pm0.009)\times10^{-5},
\end{equation}
which is consistent with the SM prediction~\cite{bib:kaon-decays} of $2.477\times10^{-5}$.

In the decay of $B$ hadrons,  $\tau$-$\ell$ universality can be tested by measurements of
the branching fractions for the exclusive decays of \BtoTauD\ and \BtoTauDstar\ expressed as ratios to the branching fractions for the exclusive
decays \BtoLD\ and \BtoLDstar, respectively. 
Systematic uncertainties due to hadronic effects largely cancel in these ratios.
A combination~\cite{bib:hfag} of the results from  the BaBar~\cite{bib:babar-btotau1,bib:babar-btotau2},
Belle~\cite{bib:belle-btotau1,bib:belle-btotau2,bib:belle-btotau3,bib:belle-btotau4,bib:belle-semilep-tagging}, and LHCb~\cite{bib:lhcb-btotau1,bib:lhcb-btotau2,bib:lhcb-btotau3} experiments yields the results:
\begin{equation}
\begin{array}{rllll}
 \displaystyle  \RD &\equiv& \displaystyle \frac{\mathcal{B}(\BtoTauD)}{\mathcal{B}(\BtoLD)} &=& 0.340 \pm 0.027 \pm 0.013, \\ \\
 \displaystyle  \RDstar &\equiv& \displaystyle \frac{\mathcal{B}(\BtoTauDstar)}{\mathcal{B}(\BtoLDstar)} &=& 0.295 \pm 0.011 \pm 0.008. \\
\end{array}
\end{equation}
These measured values exceed the SM predictions (calculated assuming LU) of:
\begin{equation}
\begin{array}{rll}
 \displaystyle  \RD &=&  0.300 \pm 0.008, \\ \\
 \displaystyle  \RDstar &=& 0.252 \pm 0.003, \\
\end{array}
\end{equation}
by 1.4$\sigma$ and 2.5$\sigma$ respectively (see~\cite{bib:hfag} and the references contained therein).
Taking into account correlations between the measurements, the combined discrepancy with regard to the SM predictions corresponds to
3.1$\sigma$~\cite{bib:hfag}.

Possible {\it Beyond the Standard Model} (BSM) explanations have been proposed for the potential violation of LU in  \RD\ and \RDstar.
A {\it Leptoquark} (LQ) that couples more strongly to the $\tau$ than to $e$ or $\mu$ could
contribute at tree level to the decays \BtoTauDandstar.
For example, in~\cite{bib:lq-btotau} a charge $2/3$ scalar LQ with $b\tau$ and $c\nu$ Yukawa couplings is able to accommodate the measured
central values of
\RD\ and \RDstar\ without introducing an unacceptable level of flavour changing neutral current processes involving the first two generations of quarks and leptons.
Interestingly, a possible link between the LEP2 measurement of \RW\ and the \BtoTauDandstar\ excess has received little attention in the literature.
We note that \RW\ might receive contributions at loop level from  a LQ that couples preferentially to the $\mathbf{\tau}$.
For example, if  $cs$ and $s\tau$ Yukawa couplings were added to the charge $2/3$ LQ scenario of~\cite{bib:lq-btotau} it could
produce an enhancement in $\mathcal{B}$(\Wtaunu).
Such loop-level contributions might naturally lead to a smaller fractional deviation from LU in \RW\ than  in \RD\ and \RDstar, but
this would depend, obviously, on the sizes of the assumed couplings.

Clearly it is important to improve on the precision of the measurements of
\RW\ [currently 2.5\%], \RD\ [currently 11\%] and \RDstar\ [currently 5\%].
The precision of the LEP2 measurement of \RW\ was dominated by the limited number of available \WW\ events.
A future high energy, high luminosity \ee\ collider, at which an improved \RW\ measurement could be performed, is likely to be
decades away. 

Measurements at hadron colliders that are sensitive to \RW\ usually rely on the identification of $\tau$ lepton decays in which the
visible final state is hadronic (\tauhad).
The current best published  hadron collider measurements typically have small statistical uncertainties, but are dominated by large systematic uncertainties that render
them uncompetitive with the LEP2 measurement of \RW.
For example, a measurement of the inclusive single \Wtauhad\ cross section in $pp$ collisions at 7~TeV by ATLAS~\cite{bib:atlas-Wtau} has a relative uncertainty of 15\%,
which is dominated by systematic uncertainties on the efficiency to trigger on and select events containing \tauhad\ candidates. 

Also of interest in this context are measurements at hadron colliders of the rate of events containing top quark-antiquark pairs (\tt), especially in the
di-lepton final state.
For the purpose of determining \RW\ events containing top quarks may be regarded as a convenient source of on-mass-shell $W$
bosons.
In the absence of non-SM decay mechanisms for the top quark the branching fraction $\mathcal{B}(\ttotaunu)$ may be reinterpreted as
the branching fraction $\mathcal{B}(\Wtaunu)$.

Relative to measurements of the inclusive single \Wtauhad\ cross section, measurements using di-leptonic \ttWWltau\ events eliminate
systematic uncertainties associated with the trigger efficiencies for \tauhad\ candidates, because single-$\ell$ ($e$ or $\mu$) triggers
can be used.
In addition, non-\tt\ backgrounds can be almost completely eliminated from the \ttWWltau\ final state by employing $b$-jet
tagging and the presence of the $\ell$ candidate.
This means that the background from misidentified hadronic jets to the \tauhad\ signature in the \ttWWltau\ final state is likely
to be significantly smaller than that in the inclusive single \Wtauhad\ final state.
However, systematic uncertainties associated with the \tauhad\ candidate (identification efficiency, background and energy scale)
still contribute directly to the measured rate of  \ttWWltau\ events.
For example, in a measurement by ATLAS~\cite{bib:atlas-top} in di-leptonic \tt\ events
the systematic uncertainty on the branching fraction $\mathcal{B}(\ttotaunu)$ is around 7.5\%.
In a measurement by CMS of the cross section for the \ttWWltau\ final state~\cite{bib:cms-top}, the systematic uncertainty is around 9.5\%; to which the combined contribution from
the identification efficiency (6.0\%), background (4.3\%), and energy scale (2.5\%) for \tauhad\ candidates was 7.8\%.
In the above-mentioned best currently published measurements from ATLAS~\cite{bib:atlas-top} and CMS~\cite{bib:cms-top} based on the  \ttWWltau\ final state the systematic uncertainties associated
with the \tauhad\ candidate are around a factor  of three greater than total uncertainty of 2.5\% on the
LEP2 measurement of \RW.

We propose here a novel, self-calibrating, ``double-ratio'' method that will allow \RW\ to be measured using top quark pair
(\ttWW) and \Ztautau\ events
at the LHC with a target precision of around 1\%, which would improve significantly upon the LEP2 measurements.
We define the ratio:
\begin{equation}
  \RbbWW \equiv \frac{N(\ttWWltau)}{N(\ttWWemu)},
  \label{eqn:RbbWW}
\end{equation}
where $N(\ttWWltau)$ and $N(\ttWWemu)$ are the numbers of observed candidate events in the \ttWWltau\ and \ttWWemu\ final states,
respectively.
We define also the ratio:
\begin{equation}
  \RZ \equiv \frac{N(\Zltau)}{N(\Zemu)},
  \label{eqn:RZ}
\end{equation}
where $N(\Zltau)$ and $N(\Zemu)$ are the numbers of observed candidate events in the \Zltau\ and \Zemu\ final states,
respectively.
We then define the double ratio:
    \begin{eqnarray}
      \RWZ &\equiv&  \frac{\RbbWW}{\RZ} \nonumber \\
   &\equiv&  \frac{N(\ttWWltau)\times N(\Zemu)}{N(\ttWWemu)\times N(\Zltau)}
  \label{eqn:RWZ}
  \end{eqnarray}
    From an experimental perspective we note that the ratios \RbbWW\ and \RZ\ are designed to have approximately the same sensitivity to systematic uncertainties
on the identification of $e$, $\mu$ and $\tauhad$ candidates, and on the efficiencies of the single-$\ell$ triggers.
Therefore, in the double ratio \RWZ\ these systematic uncertainties cancel to first order.
This cancellation is not necessarily perfect for the following reasons.
\begin{itemize}{\topsep=0pt}

\item
The distributions in  {\it transverse momentum} ($p_T$) and  {\it pseudorapidity} ($\eta$)~\cite{bib:coordinates} of the leptons ($e$, $\mu$ and $\tau$) are significantly different in the \ttWW\ and \Ztautau\ signal samples.
Systematic uncertainties on lepton identification and single-$\ell$ trigger efficiencies are not necessarily fully correlated across
all $p_T$ and \modeta\ bins.

\item
The levels of background from misidentified hadronic jets in the \tauhad\ candidates in the \ttWWltau\ and \Zltau\ samples are not necessarily identical.

\item
The level and nature of the non-\tt\ background in the \ttWWltau\ and  \ttWWemu\ samples will not necessarily be identical.
Similarly, the level and nature of the non-$Z$ boson background in the \Zltau\ and  \Zemu\ samples will not necessarily be identical.
  

\end{itemize}

From a phenomenological perspective  the double ratio \RWZ\ exploits the fact that LU has been precisely verified experimentally in
the $Z$ boson branching fractions (see equation~\ref{eqn:Z}).
Therefore, any non-SM effects should affect \RWZ\ primarily through \RbbWW, which is sensitive to \RW.

The rest of this paper is organised as follows.
In section~\ref{sec:MC} we describe a {\it Monte Carlo} (MC) study of the proposed analysis method employing a simple parameterised
simulation of detector performance.
In section~\ref{sec:conclusions} we present our summary and conclusions.


\section{A Monte Carlo study of the proposed double-ratio analysis method}    
\label{sec:MC}

Our study uses particle-level MC events for the various physics processes of relevance.
In section~\ref{sec:MCdetector} we describe briefly the simple parameterised
simulation of detector performance we use in the  study of the proposed double-ratio analysis method.
We describe also the variations in detector performance we consider in the study of experimental systematic uncertainties.
Further details are given in the Appendix.
In section~\ref{sec:MCgenerators} we describe the MC generators used and the potential sources of phenomenological systematic uncertainties
that we have considered in our study.
In section~\ref{sec:MCcuts} we describe the candidate event selection criteria we employ for the four signal candidate event samples
used in the  double-ratio method.
In the selection of \Ztautau\ candidates we propose a novel selection variable, \mstar, that improves the discrimination power
against the dominant backgrounds, such as \tt, diboson production, as well as events containing a leptonically decaying vector boson plus a
QCD jet that is misidentified as a \tauhad\ candidate ($W$+jet).
In section~\ref{sec:MCsamples} we evaluate the size and composition of the four selected candidate event samples.
The expected numbers of events are given for an integrated luminosity at $\sqrt{s}$~=~13~TeV of \intL~=~140~\invfb, which corresponds
approximately to that available for physics analysis in ATLAS and CMS at the end of LHC run~2~\cite{bib:atlas-140,bib:cms-140}.
In section~\ref{sec:MCsensitivity} we evaluate the sensitivity of the measured double ratio \RWZ\ to the physical quantity of
interest \RW.
In section~\ref{sec:MCsyst} we present the effect of systematic uncertainties on the ratios \RbbWW, \RZ, \RWZ, and \RW.
We thus demonstrate that in the double-ratio \RWZ\  there is a high degree of cancellation in the experimental systematic uncertainties that have dominated previous
related analyses at hadron colliders.
We evaluate the residual systematic uncertainties on \RWZ\ arising from the effects discussed in sections~\ref{sec:MCdetector} and~\ref{sec:MCgenerators}.
In section~\ref{sec:MClimitations} we discuss some limitations of this simplified study and consider some factors that will need to
be taken into account in an analysis that uses detailed simulations of a specific LHC detector and is applied to the experimental data.

\subsection{Simple parameterised Monte Carlo simulation of detector performance}    
\label{sec:MCdetector}

In this section we describe briefly the simple parameterised
simulation of detector performance we use to study the proposed double-ratio analysis method.
We describe also the variations in detector performance we consider in the study of systematic uncertainties.

Clearly, our aim here is not to produce a completely accurate simulation of the data from either the ATLAS or CMS detectors.
Nevertheless, we base our parameterisations of the detection of leptons and jets on published measurements of LHC detector performance and their associated uncertainties~\cite{bib:atlas-muonID}--\cite{bib:pile-up}.
This approach enables us to demonstrate some of the principle benefits of the proposed double-ratio method and allows us to
investigate within a simple and controlled framework the principal sources
of residual systematic uncertainty to which the method is sensitive.

The efficiencies associated with the reconstruction, identification, and triggering of high $p_T$, isolated lepton candidates are
typically determined by the ATLAS and CMS collaborations using ``tag and probe'' measurements on \Zemutau\ events in both MC
simulations and the real data.
Systematic uncertainties are usually quoted on the ``scale factors'' employed to correct MC simulations to provide an accurate
description of the real data. 

As noted in section~\ref{sec:intro}, the ratios \RbbWW\ and \RZ\ are designed to have approximately the same sensitivity to systematic uncertainties
on the identification of $e$, $\mu$ and $\tauhad$ candidates, and on the efficiencies of the single-$\ell$ triggers.
In the double ratio \RWZ, these systematic uncertainties cancel to first order.
Therefore, in our study we give particular attention to the $p_T$ and $\eta$ dependence of efficiencies and to any potential $p_T$ and $\eta$ dependence in the associated systematic uncertainties.
In general we expect algorithms that are designed to have $p_T$- and $\eta$-independent efficiencies to have smaller $p_T$- and $\eta$-dependent systematic uncertainties.

\begin{figure}[hbtp]
\centering
\includegraphics[width=0.5\textwidth]{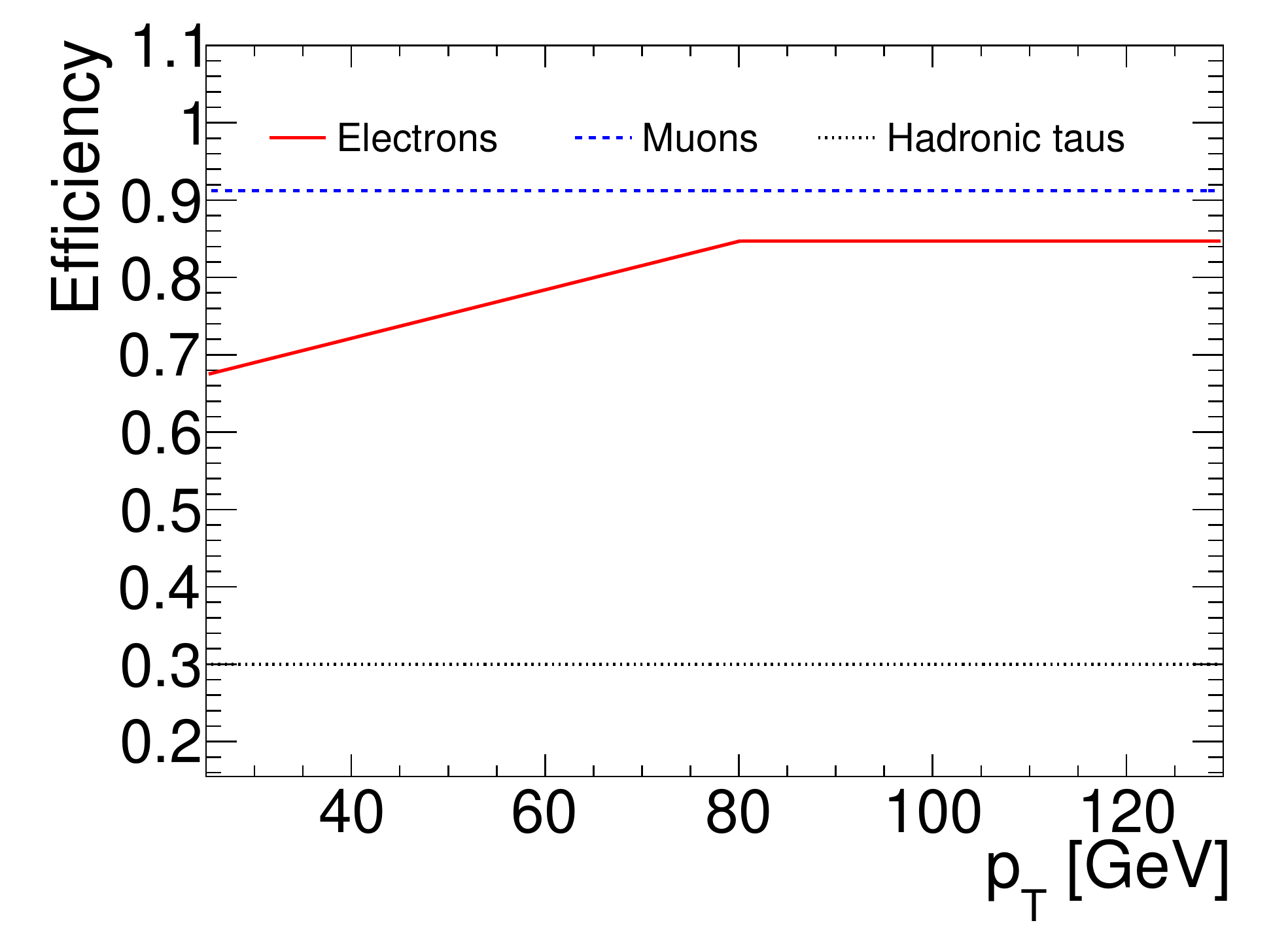}
\caption{
  The overall efficiencies assumed for the reconstruction,
identification, and isolation criteria for prompt $e, \mu$, and \tauhad\ candidates as a function of~$p_T$.
}
\label{fig:efficiency}
\end{figure}

\begin{figure}[hbtp]
\centering
\includegraphics[width=0.5\textwidth]{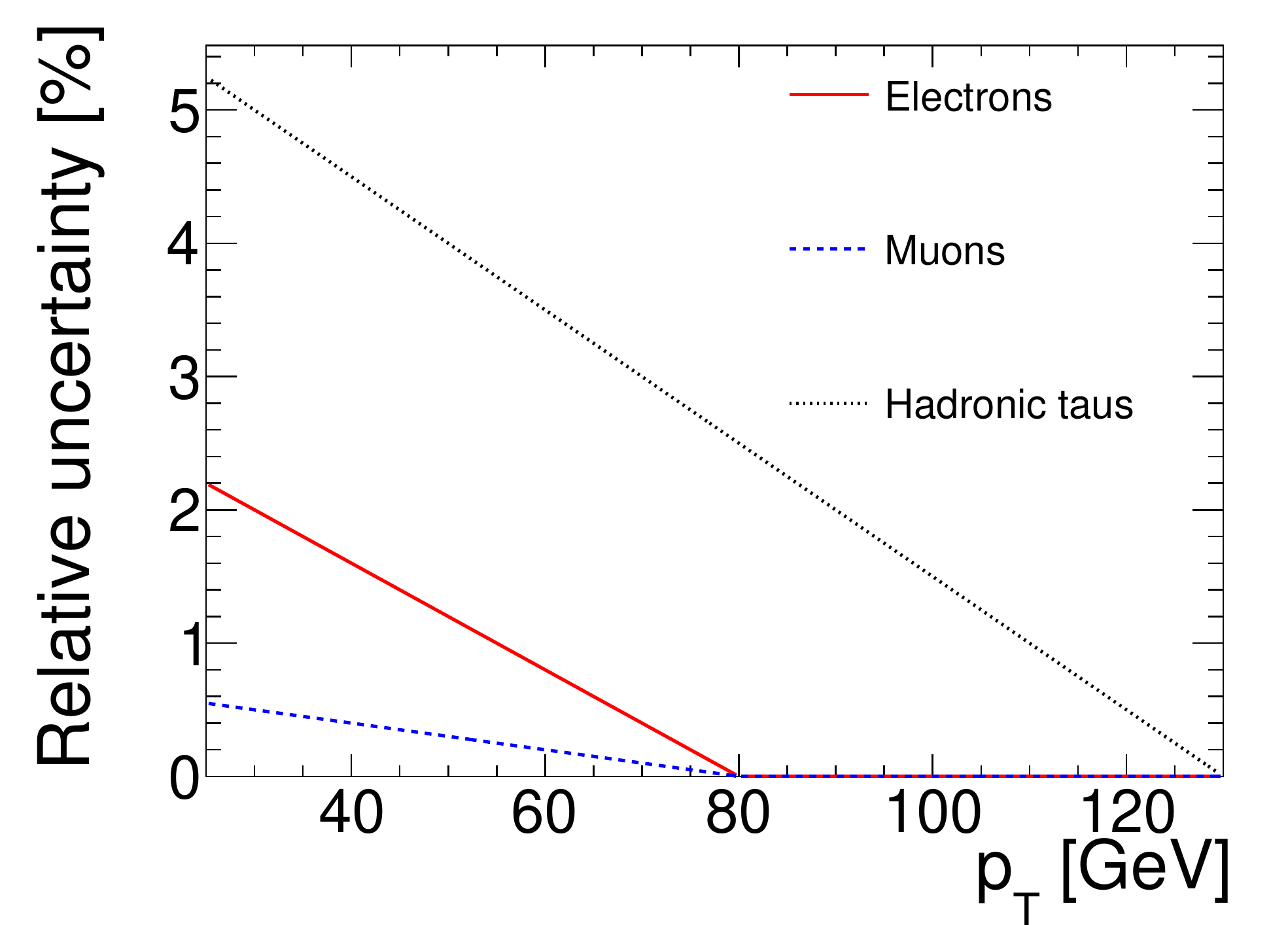}
\caption{
  The $p_T$-dependent systematic uncertainties on the efficiencies we assume for the reconstruction,
identification, and isolation criteria for prompt $e, \mu$, and \tauhad\ candidates.
}
\label{fig:efficiency-syst}
\end{figure}

Figure~\ref{fig:efficiency} shows as a function of $p_T$ the overall efficiencies we assume for the reconstruction,
identification, and isolation criteria for prompt $e, \mu$, and \tauhad\ candidates.
Table~\ref{tab:systematics} gives a summary of the sources of systematic uncertainty on the reconstruction of leptons and jets
considered in this study, stating separately the
assumed size of the
$p_T$/$\eta$-independent and $p_T$/$\eta$-dependent systematic uncertainties.
Figure~\ref{fig:efficiency-syst} shows the $p_T$-dependent systematic uncertainties on the efficiencies we assume for the reconstruction,
identification, and isolation criteria for prompt $e, \mu$, and \tauhad\ candidates.
\begin{table*}[hbtp]
\centering
\begin{tabular}{|c|c|c|} 
\hline
Source of & \multicolumn{2}{|c|}{Size of systematic uncertainty (in \%, unless stated otherwise)}  \\
\cline{2-3}
systematic uncertainty & $p_T$/$\eta$-independent & $p_T$/$\eta$-dependent  \\
\hline
 &    &     \\
Muon ID \& isolation efficiency & 2.0  & $\displaystyle 0.5\left(\frac{80-p_T}{50}\right)$, 0 for $p_T>80\,\mathrm{GeV}$  \\
 &    &     \\
Single-muon trigger efficiency & 1.0  & 1.0 in endcap, 0 in barrel  \\
Muon $p_T$ resolution &  0.15  & 0.15 in endcap, 0 in barrel   \\
Muon $p_T$ scale &  0.2  & 0.2 in endcap, 0 in barrel   \\
 &    &     \\
Electron ID \& isolation efficiency & 2.0  & $\displaystyle 2.0\left(\frac{80-p_T}{50}\right)$, 0 for $p_T>80\,\mathrm{GeV}$  \\
 &    &     \\
Single-electron trigger efficiency & 1.0  & 1~GeV change in $p_T$ threshold  \\
 &    &     \\
Electron $p_T$ resolution & vary constant term in  &  vary resolution in endcap only \\
 & $\frac{\sigma_{p_T}}{p_T}$ by  $\pm 0.002$   &  \\
 &    &     \\
Electron $p_T$ scale &  0.2  & 0.2 in endcap, 0 in barrel   \\
 &    &     \\
\tauhad\ efficiency         &  5  &  $\displaystyle 5\left(\frac{130-p_T}{100}\right)$, 0 for $p_T>130\,\mathrm{GeV}$   \\
 &    &     \\
\tauhad\ $p_T$ scale        &  1  &  1  in endcap, 0 in barrel  \\
 &    &     \\
jet energy scale & 1   &  ---   \\
$b$-tag efficiency for $b$-, $c$-, light-jets & 1.5, 4, 10   & ---    \\
 &    &     \\
Misidentification rates for \tauhad\ & 10   & ---    \\
MJ background in  \Zltau\ sample & 5   & ---   \\
 &    &     \\
\hline
\end{tabular}
\caption{Sources of experimental systematic uncertainty, together with the size (in percent, unless stated otherwise) of the
  $p_T$/$\eta$-independent and $p_T$/$\eta$-dependent systematic uncertainties considered in this study.
  In all expressions $p_T$ is given in units of GeV.
  We define the endcap region by $|\eta|>1.0$.
  See the Appendix for further detail and justification of these choices.
}
\label{tab:systematics}
\end{table*}

Jets with \modeta~$<$~2.5 are flagged as $b$-tagged with probabilities that depend on their flavour at truth level as follows: truth $b$ 85\%, truth $c$ 40\%, truth
light quark or gluon 1\%.
These tag probabilities are approximately independent of $p_T$ and \modeta.

A detailed description of our choices of lepton and jet identification algorithms to simulate and experimental systematic uncertainties to
consider, and the reasoning behind these choices, are given in the Appendix.

\subsection{Monte Carlo generators and phenomenological systematic uncertainties}
\label{sec:MCgenerators}

Events containing $W$ bosons, $Z$ bosons, top quark pairs, and the EW production of single top quarks are generated using {\sc
  POWHEG BOX}~\cite{bib:powheg}, interfaced to {\sc PYTHIA}\cite{bib:pythia} for the simulation of parton showering and
fragmentation.
We thus ensure a consistent treatment of tau lepton decay in the principle sources of candidate events.
EW diboson events are generated using {\sc SHERPA}~\cite{bib:sherpa}.

Significant sources of  background in the selected \Ztautau\ samples arise from EW diboson and \tt\  production.
Cross sections at $\sqrt{s}$~=~13~TeV have been measured for  EW diboson~\cite{bib:atlas-ww} and \tt~\cite{bib:cms-tt}  production.
Our estimates of the fractional composition of the selected \Ztautau\ samples are not sensitive to systematic uncertainties on
the integrated luminosity or on the predicted absolute cross sections for $Z$ boson, \tt, or EW diboson production.
They are, however, sensitive  to systematic uncertainties on
the ratio of the cross sections for \tt\ and EW diboson production to that for $Z$ bosons.
We assume an uncertainty of 3\% on both of these cross section ratios~\cite{bib:ww-tt-z-ratios}.

In \Ztautau\ events the momentum distribution of electrons and muons produced in $\tau$ decay is softer than that for the visible \tauhad\ systems.
Changes in the distribution of the transverse momentum of the produced $Z$ bosons (\Zpt) can, therefore, affect the relative
acceptance for \Zltau\ and \Zemu\ candidate events.
Measurements of \Zpt\ have been made at  $\sqrt{s}$~=~8~TeV by ATLAS~\cite{bib:atlas-zptphistar}.
In order to evaluate systematic uncertainties arising from \Zpt\ we increase
the weight of events satisfying $50\ \mathrm{GeV} < \Zpt  < 150$~GeV by 0.5\% and
the weight of events satisfying $\Zpt > 150$~GeV by 1.0\%~\cite{bib:atlas-zptsyst}.

In \ttWWltau\ events, the majority of observed electrons and muons are from direct $W$ boson decays and therefore have a distribution in
$p_T$ that is much harder than that for the visible \tauhad\ systems.
The relative acceptance for \ttWWltau\ and \ttWWemu\ final states may be sensitive to the details of the modelling of \tt\ production.
We investigate this sensitivity by using an alternative \tt\ generator-level sample in which the QCD factorisation and
renormalisation scales are changed by a factor of two relative to the default values.
In addition, we evaluate systematic uncertainties arising from simulating the distribution of the mass of the \tt\ system, \mtt,\ by increasing the weights of events by an amount that varies
linearly between 1\% at \mtt\ = 500~GeV to 10\% at \mtt\ = 1000~GeV~\cite{bib:atlas-top-diff-cross-section}.

Table~\ref{tab:pheno-systematics} gives a summary of the sources of phenomenological  systematic uncertainty considered in this study.
\begin{table*}[tphb]
\centering
\begin{tabular}{|c|c|} 
\hline
Source of systematic uncertainty & Size of fractional systematic uncertainty (in \%)  \\
\hline
Ratio of diboson and $Z$ boson cross sections    &  3   \\
Ratio of \tt\ and $Z$ boson cross sections    &  3   \\
\Zpt\ reweighting    &  0.5 for $50 < \Zpt  < 150$ \\
                     &  1.0 for $\Zpt > 150$   \\
\mtt\ reweighting & $10 + 0.018(m(t\bar{t})-1000)$  \\
\tt\ modelling & Tested with alternative sample \\
\hline
\end{tabular}
\caption{Sources of phenomenological systematic uncertainty considered in this study (given in percent).
  In all expressions, $p_T$ and $m(t\bar{t})$ are given in units of GeV.
  See the text for further details.
}
\label{tab:pheno-systematics}
\end{table*}

\subsection{Candidate event selection criteria}    
\label{sec:MCcuts}

Candidate electrons and muons are considered in the analysis if after simulation of resolution and momentum scale
they satisfy  $p_T > 27$~GeV.
Candidate \tauhad\ and hadronic jets are considered if after simulation of resolution and momentum scale
they satisfy $p_T > 25$~GeV.
Hadronic jets must satisfy \modeta~$<$~4.5.
Candidate electrons, muons, \tauhad, and $b$-tagged jets must satisfy \modeta~$<$~2.5.

\subsubsection{Candidate event selection criteria on leptons}    
\label{sec:MCleptoncuts}

In order to maximise the cancellation of systematic uncertainties between \ttWW\ and \Ztautau\ event samples the same candidate event
selection criteria on leptons are applied in the two event classes.

\medskip
\noindent
Candidate  \etau\  events are required to contain:
\begin{itemize}{\topsep=0pt}

\item
Exactly one $e$ candidate.

\item
Exactly one \tauhad\ candidate of opposite sign to the $e$ candidate. 

\item
No $\mu$ candidates.

\item
The $e$ candidate must fire the single-$e$ trigger.

\end{itemize}

\noindent
Candidate  \mutau\  events are required to contain:
\begin{itemize}{\topsep=0pt}

\item
Exactly one $\mu$ candidate.

\item
Exactly one \tauhad\ candidate of opposite sign to the $\mu$ candidate. 

\item
No $e$ candidates.

\item
The $\mu$ candidate must fire the single-$\mu$ trigger.

\end{itemize}

\noindent
Candidate  \emu\  events are required to contain:
\begin{itemize}{\topsep=0pt}

\item
Exactly one $e$ candidate.

\item
Exactly one $\mu$ candidate of opposite sign to the $e$ candidate.

\item
The $e$ candidate must fire the single-$e$ trigger, and/or the $\mu$ candidate must fire the single-$\mu$ trigger.

\end{itemize}

\subsubsection{\tt\ candidate event selection criteria}    

In addition to the relevant criteria on leptons given in section~\ref{sec:MCleptoncuts} above, all candidate \ttWW\ events
(\ttWWltau\ as well as \ttWWemu ) are required to contain exactly two $b$-tagged jets.
Because the same criterion is applied in the selection of the numerator
\ttWWltau\ and denominator \ttWWemu\ events, we expect the ratio \RbbWW\ to be largely insensitive to systematic uncertainties associated with jet reconstruction, JES, JER, and $b$-tagging.
Requiring two $b$-tagged jets reduces the background in the selected \ttWW\ event samples from non-\tt\ sources; it also reduces the
background from \tt\ events in which a $b$-quark jet is misidentified as a prompt $e, \mu$, or \tauhad.

\begin{figure}[hbtp]
\centering
\includegraphics[width=0.45\textwidth]{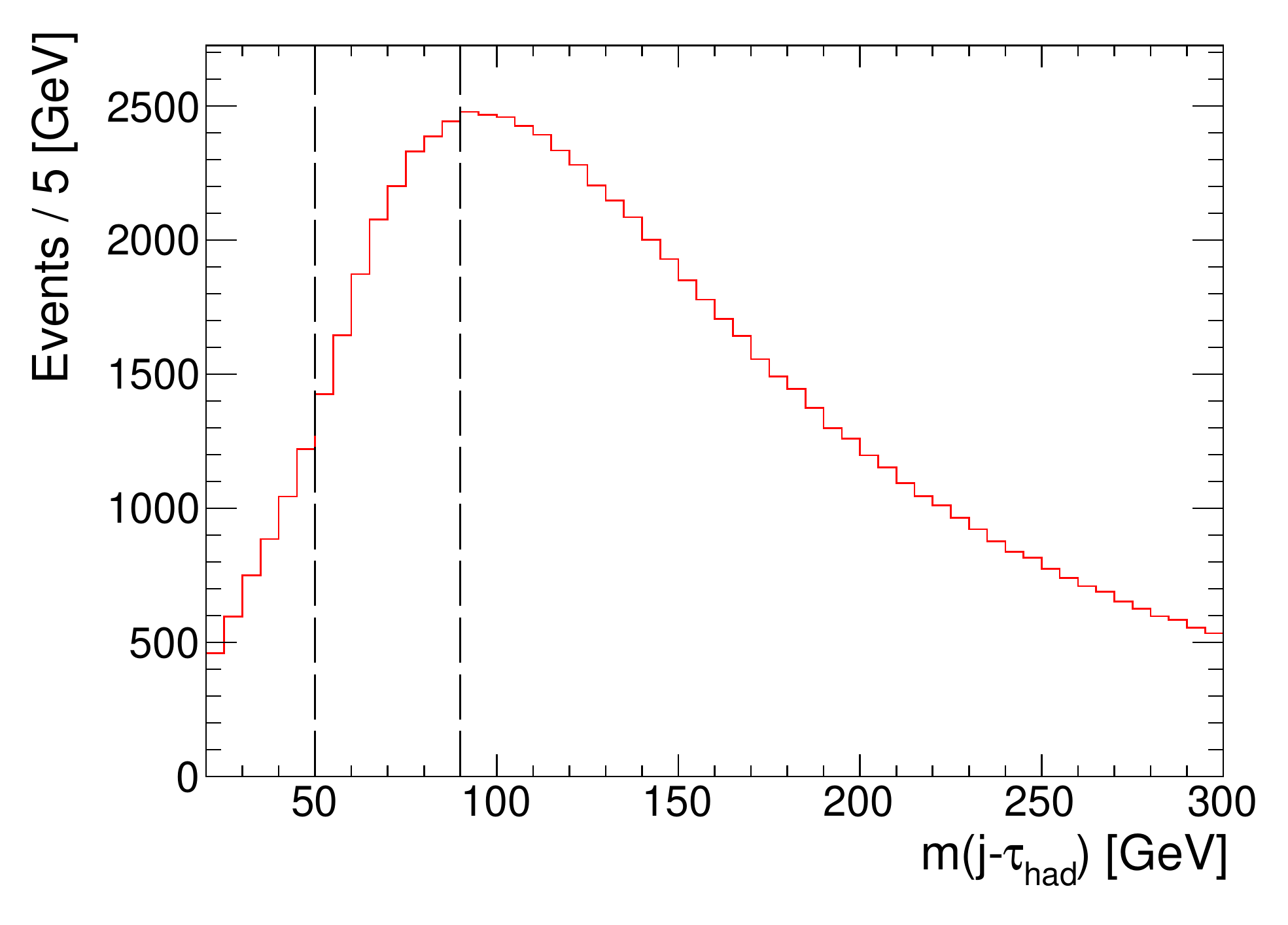}
\includegraphics[width=0.45\textwidth]{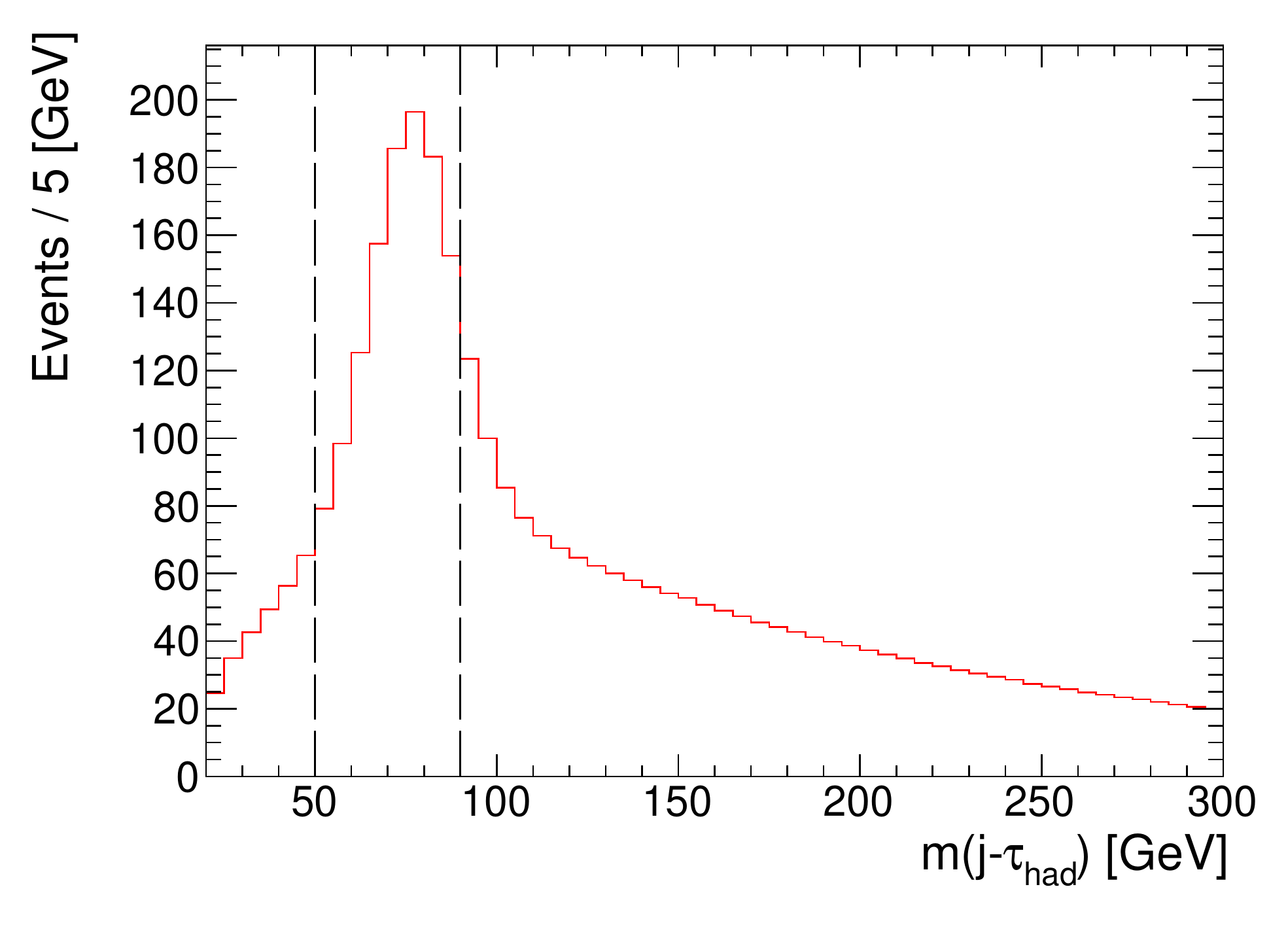}
\caption{
  The distribution of \mtaujet\ in the \ttWWltau\ candidate event sample. 
  The event selection requirement on this quantity has not been applied.
  The upper plot shows events in which the \tauhad\ candidate originates from a genuine $\tau$ decay and the lower plot
  shows events in which the \tauhad\ candidate originates from a misidentified hadronic jet.
}
\label{fig:mass-tau-jet}
\end{figure}

Candidate \ttWWltau\ events are rejected if the invariant mass of the \tauhad\ candidate and the highest $p_T$
non-$b$-tagged jet, \mtaujet, satisfies 50~GeV~$<$ \mtaujet\ $<$ 90~GeV.
This criterion reduces the background from lepton+jet \tt\ events in which the hadronically
decaying $W$ boson produces two reconstructed jets, one of which is misidentified as a \tauhad\ candidate.
The effectiveness of this criterion is illustrated by Figure~\ref{fig:mass-tau-jet}, which
 shows in the \ttWWltau\ candidate event sample the distribution of \mtaujet, having applied all other \ttWWltau\ event selection criteria.
  The upper plot shows events in which the \tauhad\ candidate originates from a genuine $\tau$ decay and the lower plot
  shows events in which the \tauhad\ candidate originates from a misidentified hadronic jet.
  A clear peak at around the mass of the $W$ boson can be seen in the lower plot.
  In addition to helping reject background from misidentified hadronic jets, the distributions in Figure~\ref{fig:mass-tau-jet} offer the possibility to make a
  data-driven estimate of the background in the  \tauhad\ candidate sample.
This will be useful in reducing the systematic uncertainty on the background yield.

\subsubsection{\Ztautau\ candidate event selection criteria}

In addition to the relevant criteria on leptons given in section~\ref{sec:MCleptoncuts} above, all candidate \Ztautau\  events  (\Zltau\ as well as \Zemu ) are required to satisfy the following criteria:

\begin{itemize}{\topsep=0pt}

\item
Events should contain no $b$-tagged jets.

\item
50~GeV~$<\mstar <$~100~GeV.  

\item
$\sumcos >$~$-0.1$.  


\item
$\at < 60$~GeV.


\end{itemize}

\begin{figure*}[hbtp]
\centering
\includegraphics[width=0.45\textwidth]{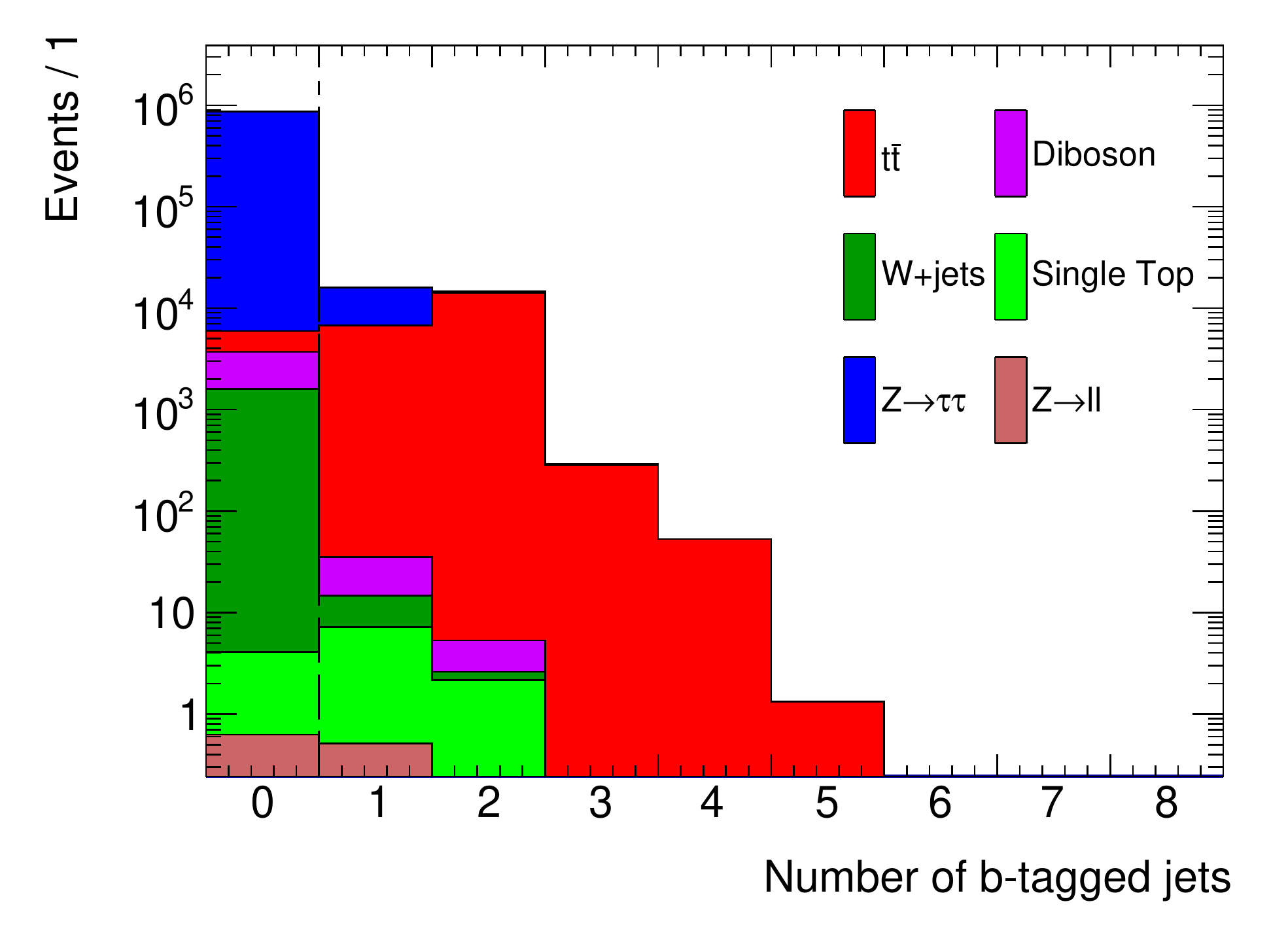}
\includegraphics[width=0.45\textwidth]{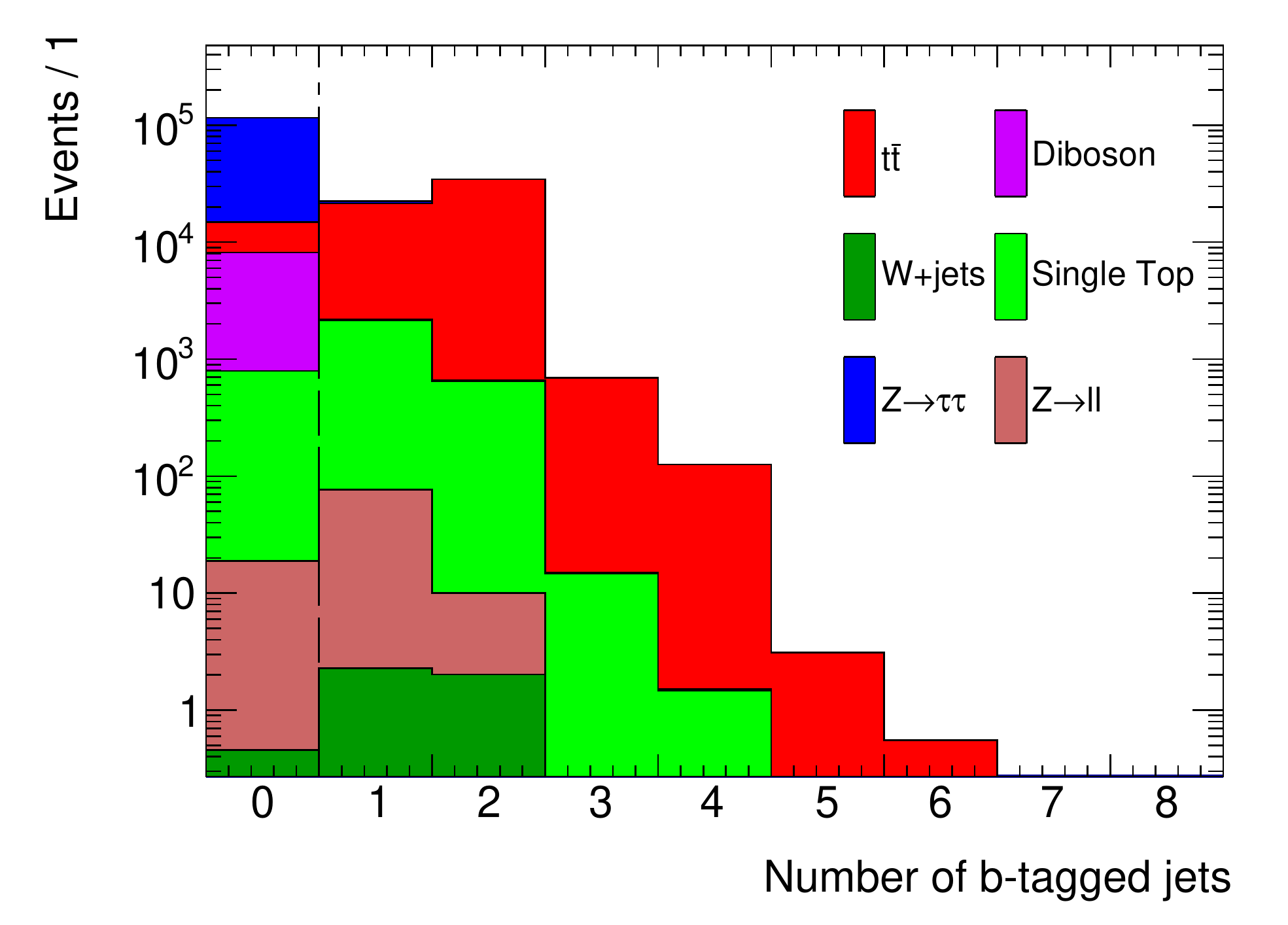}
\includegraphics[width=0.45\textwidth]{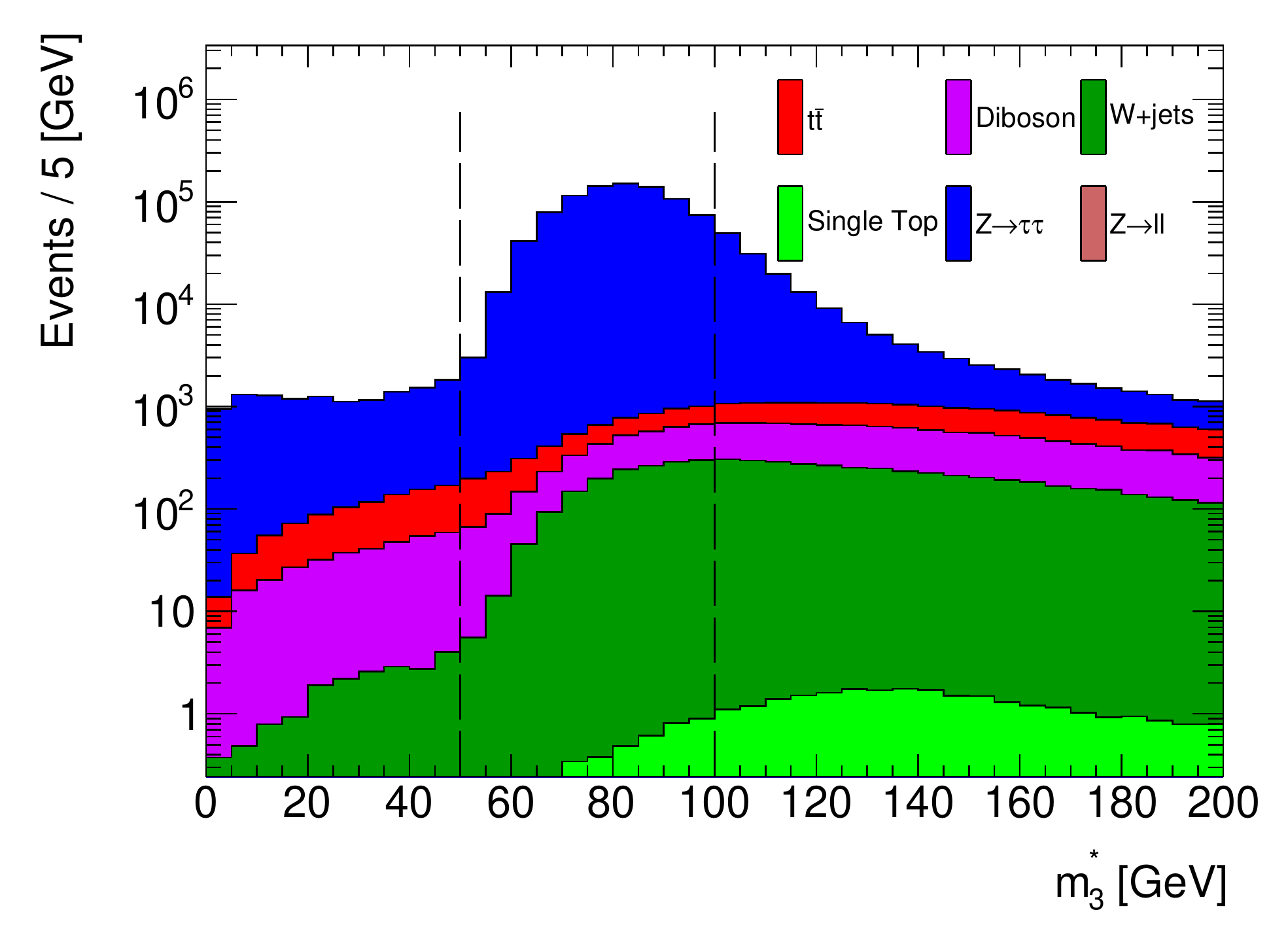}
\includegraphics[width=0.45\textwidth]{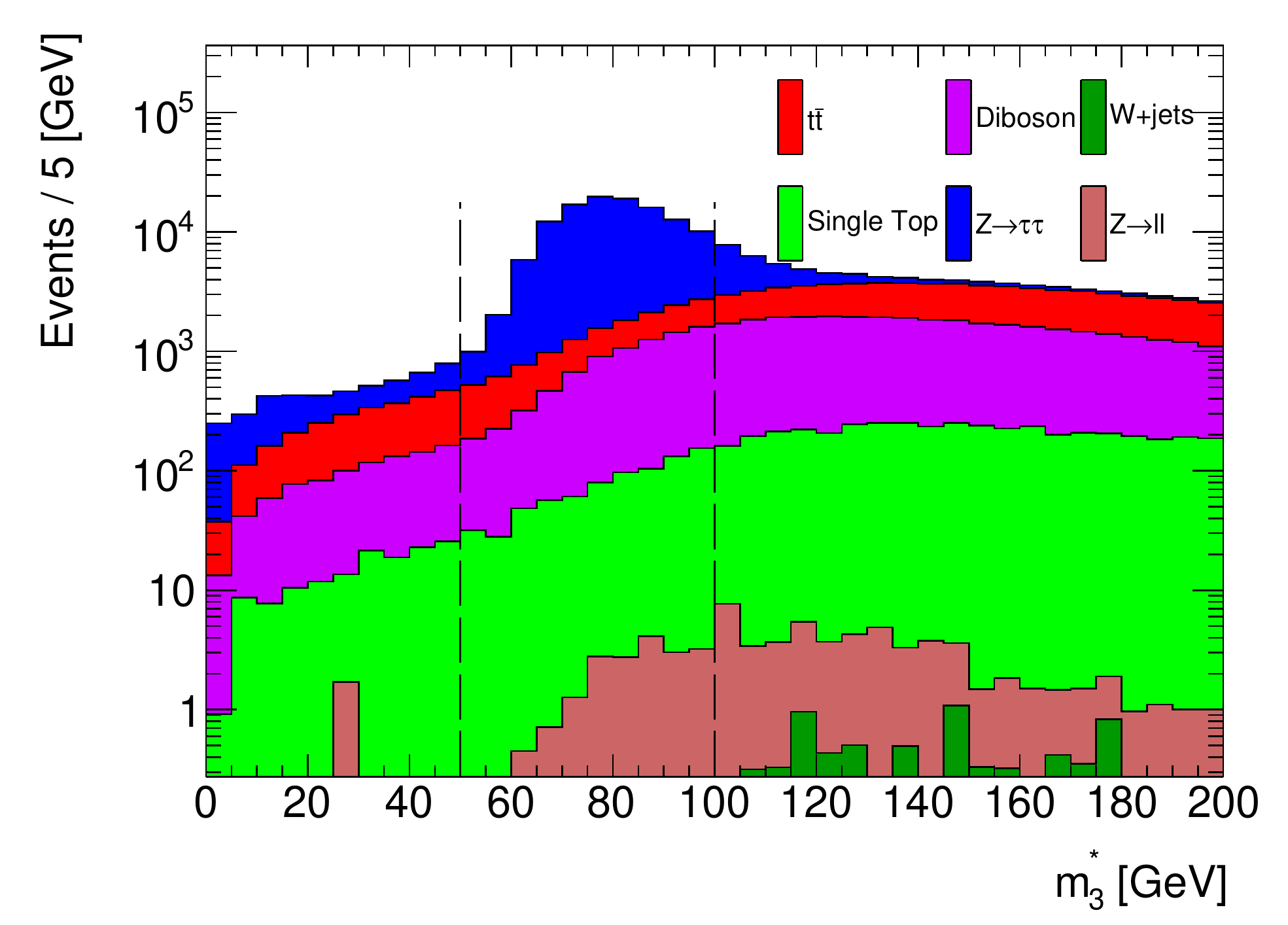}
\includegraphics[width=0.45\textwidth]{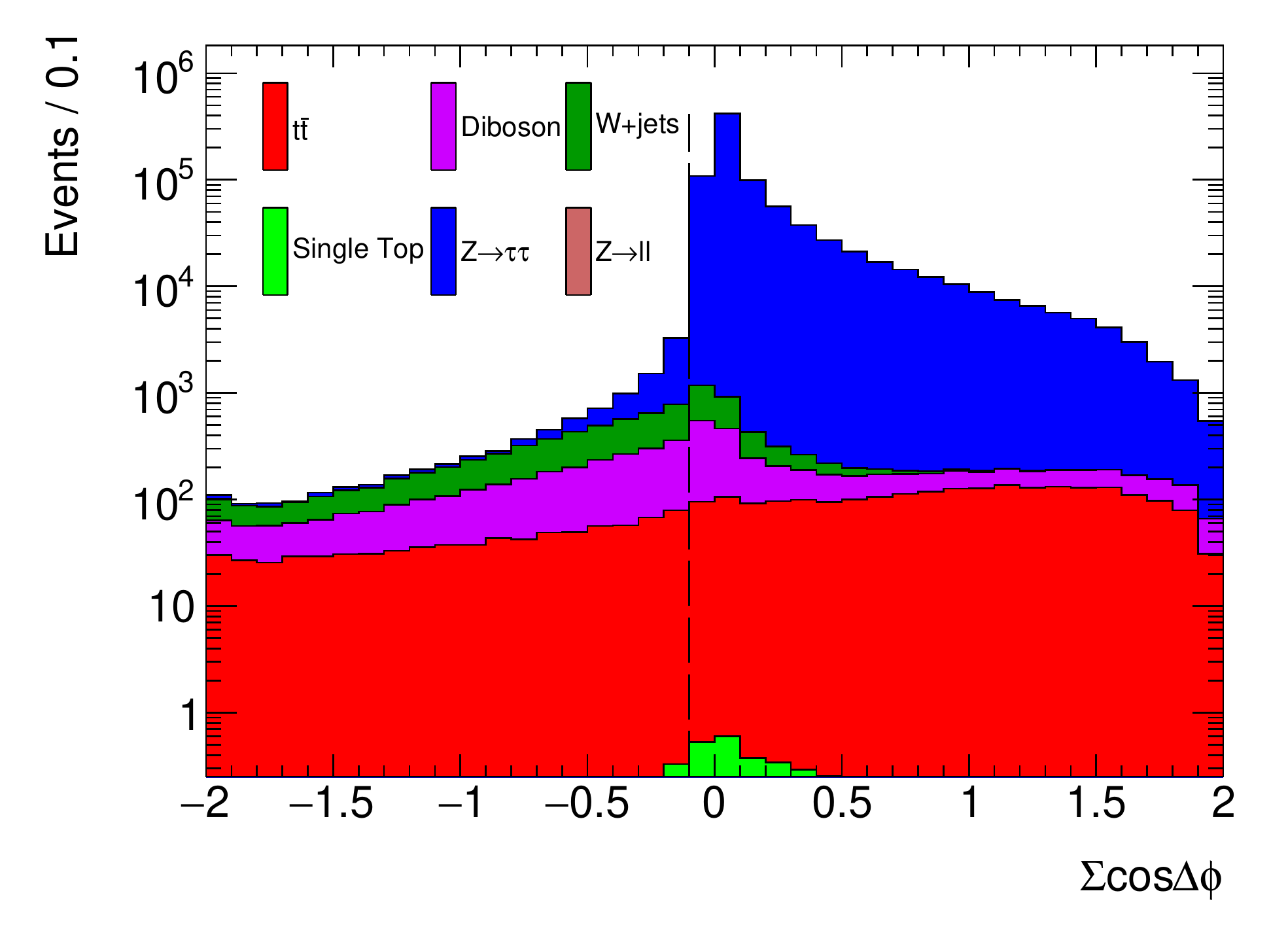}
\includegraphics[width=0.45\textwidth]{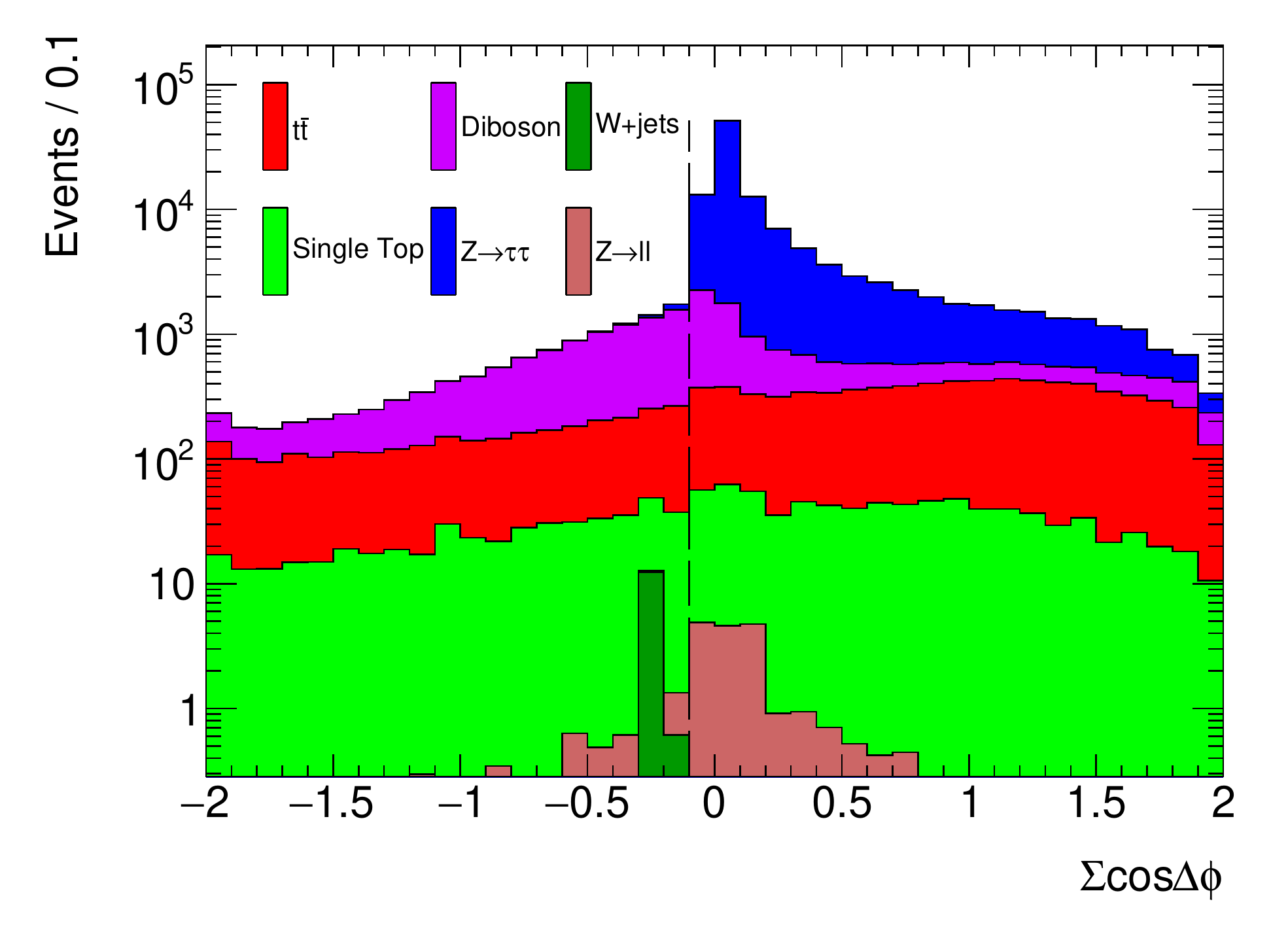}
\includegraphics[width=0.45\textwidth]{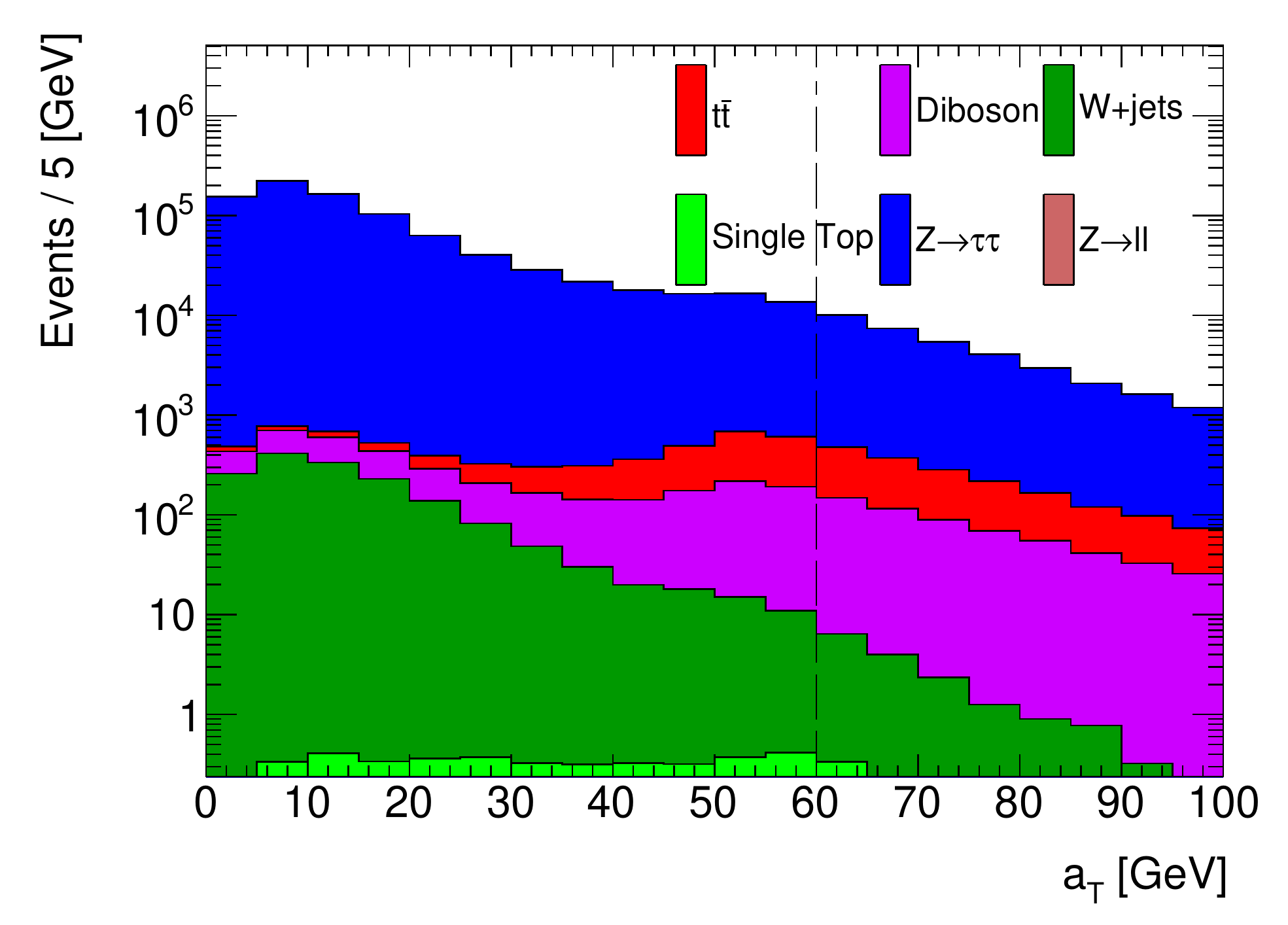}
\includegraphics[width=0.45\textwidth]{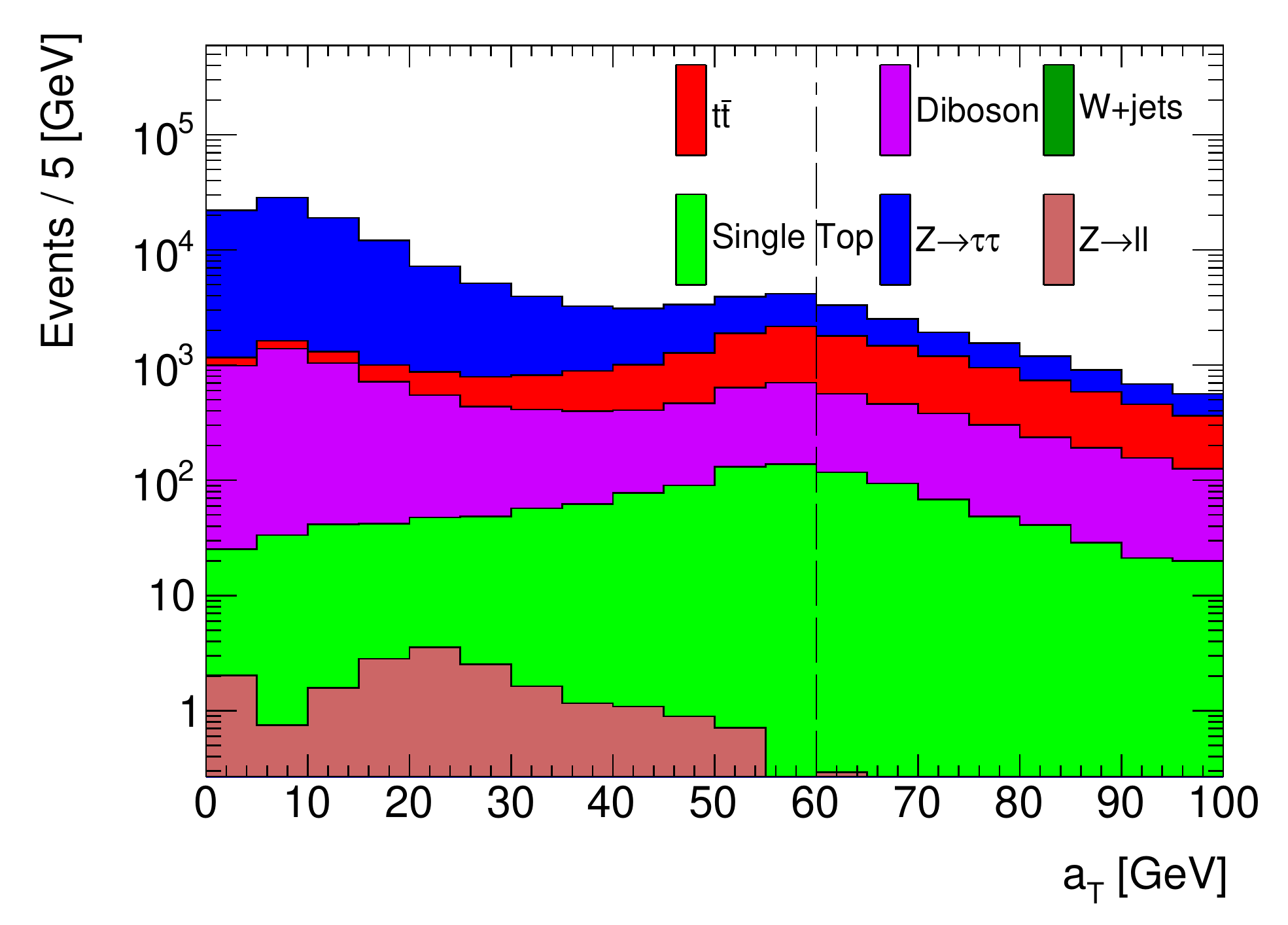}
\caption{Distributions of  variables in the selection of \Ztautau\ candidate events having applied all other selection criteria.
  The left column shows \Zltau\ candidate events.
  The right column shows \Zemu\ candidate events.
  The selection criteria are indicated by vertical lines.
}
\label{fig:Ztautau_selection}
\end{figure*}

The relevant variables are defined below.
Distributions of each variable having applied all other selection criteria are shown in Figure~\ref{fig:Ztautau_selection} for
\Zltau\ and \Zemu.

As expected, the rejected event samples containing $b$-tagged jets are dominated by \tt.
Because the same criterion on $b$-tagged jets is applied in the selection of the numerator
\Zltau\ and denominator \Zemu\ events, we expect the ratio \RZ\ to be largely insensitive to systematic uncertainties associated with jet reconstruction, JES, JER, and $b$-tagging.
Requiring no $b$-tagged jets reduces the background in the selected \Ztautau\ event samples from \tt\  and also $W$ boson plus heavy
flavour production  in which a $b$-quark jet is misidentified as a prompt $e$, $\mu$, or \tauhad.
It can be seen that the other selection criteria reject a large fraction of the remaining background, principally from $W$+jet production (in the \Zltau\ sample) and from \tt\ and diboson production (in the \Zemu\ sample).

\begin{figure}[hbtp]
\includegraphics[width=0.45\textwidth]{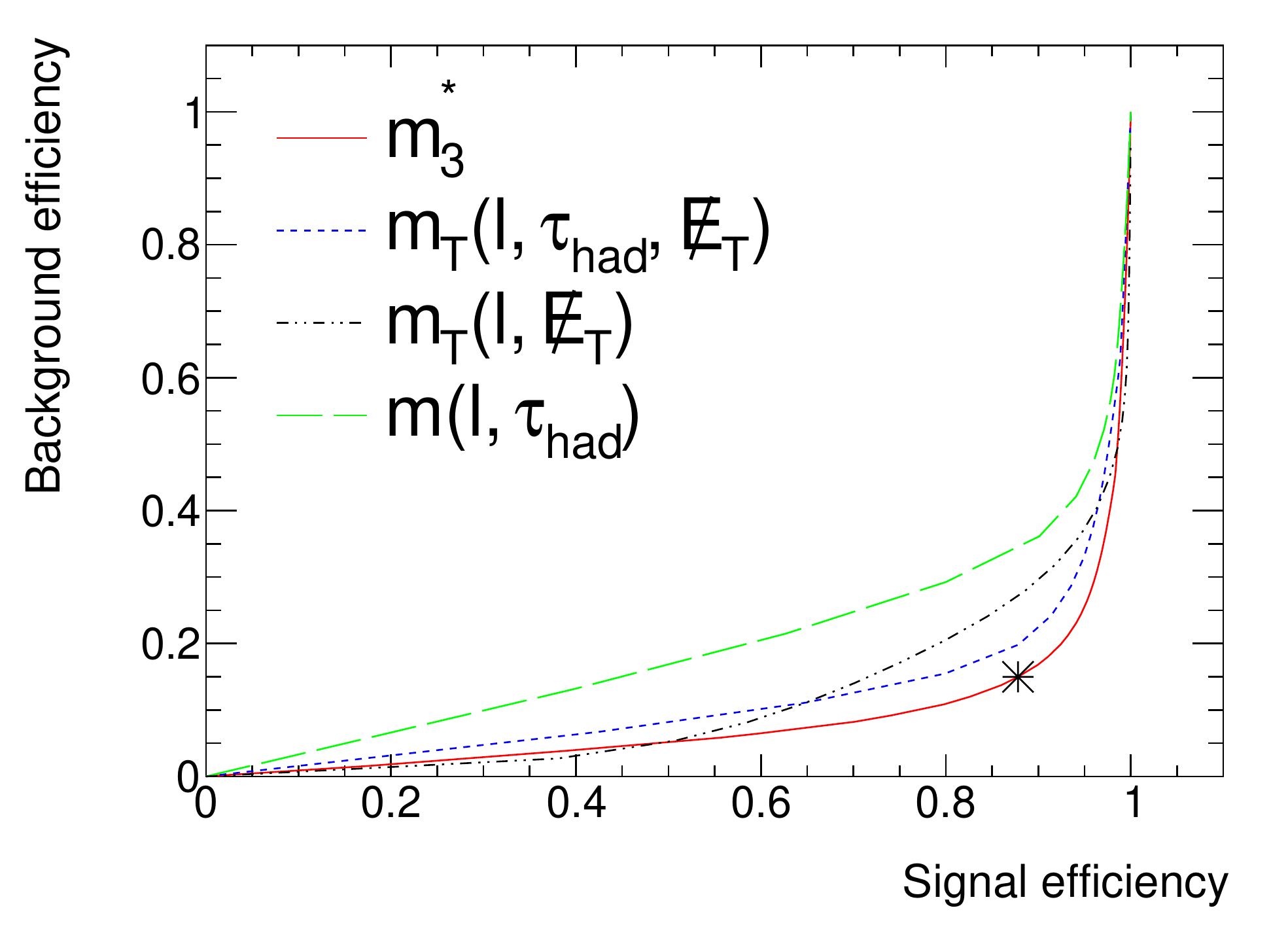}
\caption{A comparison of the signal and background efficiencies for cuts on different mass variables in the \Zltau\ candidate event sample.
The variable $m(\ell, \tauhad)$ is defined as the visible mass of the lepton and \tauhad\ candidate.
The star indicates the position of the proposed cut on \mstar, which outperforms the other variables over the range of interest.}
\label{fig:ROC}
\end{figure}

We propose here a novel selection variable, \mstar:
\begin{equation}
  \mstar \equiv \frac{m_T(\ell, \tauhad, \etmiss)}{\sts}.
\end{equation}
Here  $\slashed{E}_T$ is the missing transverse
momentum and
the transverse mass, $m_T(\ell, \tauhad, \etmiss)$, of the 3-body system of $\ell$, \tauhad, and  $\slashed{E}_T$
 may be defined by:
$$  m_T(\ell, \tauhad, \etmiss)^2 = \qquad\qquad\qquad\qquad\qquad\qquad\qquad\qquad $$
\begin{equation}
\\ \qquad m_T(\ell, \tauhad)^2 + m_T(\ell, \etmiss)^2 + m_T(\tauhad, \etmiss)^2 
\end{equation}
$\theta^*_\eta $ is an approximation to the scattering angle
of the leptons relative to the beam direction in the dilepton rest frame.
This variable is defined~\cite{bib:sa-phistar} solely using the measured track directions by:
\begin{equation}
\cos(\theta^{*}_{\eta}) \equiv \tanh\left(\frac{\eta^--\eta^+}{2}\right),
\end{equation}
where $\eta^-$ and $\eta^+$ are the pseudorapidities of the negatively
and positively charge lepton, respectively.
The division by \sts\ in the definition of \mstar\ takes into account the relative longitudinal motion of the two leptons and, therefore, \mstar\ is a
more closely correlated with the \tautau\ mass than is $m_T(\ell, \tauhad, \etmiss)$.
In the selection of \Ztautau\ candidates this improves the discrimination power
against the dominant backgrounds (such as $W$+jet and diboson events).
A comparison of the performance of \mstar\ with other similar discriminating variables is shown in Figure~\ref{fig:ROC}.

The variable \sumcos~\cite{bib:sumcos}:
\begin{equation}
  \sumcos \equiv \cos(\Delta\phi(\ell,\etmiss)) + \cos(\Delta\phi(\tauhad,\etmiss)),
\end{equation}
discriminates against background events containing leptonically decaying $W$ bosons.
Here $\Delta\phi(\ell,\etmiss)$ is the azimuthal angle between the $\ell$ and the \etmiss\ and $\Delta\phi(\tauhad,\etmiss)$ is the azimuthal angle between the \tauhad\ and the \etmiss.

The variable \at~\cite{bib:at} corresponds to the component
of the $p_T$ of the dilepton system that is transverse to the dilepton thrust axis.
This variable is well suited to the study of \tautau\ final states, because it is less sensitive to any imbalance in the transverse
momenta of the neutrinos produced in the tau decays than is \al~\cite{bib:at},  the component
of the dilepton $p_T$ that is longitudinal to the dilepton thrust axis.
The variable \at\ discriminates against background events containing leptonically decaying $W$ bosons.

\begin{figure*}[hbtp]
\centering
\includegraphics[width=0.4\textwidth]{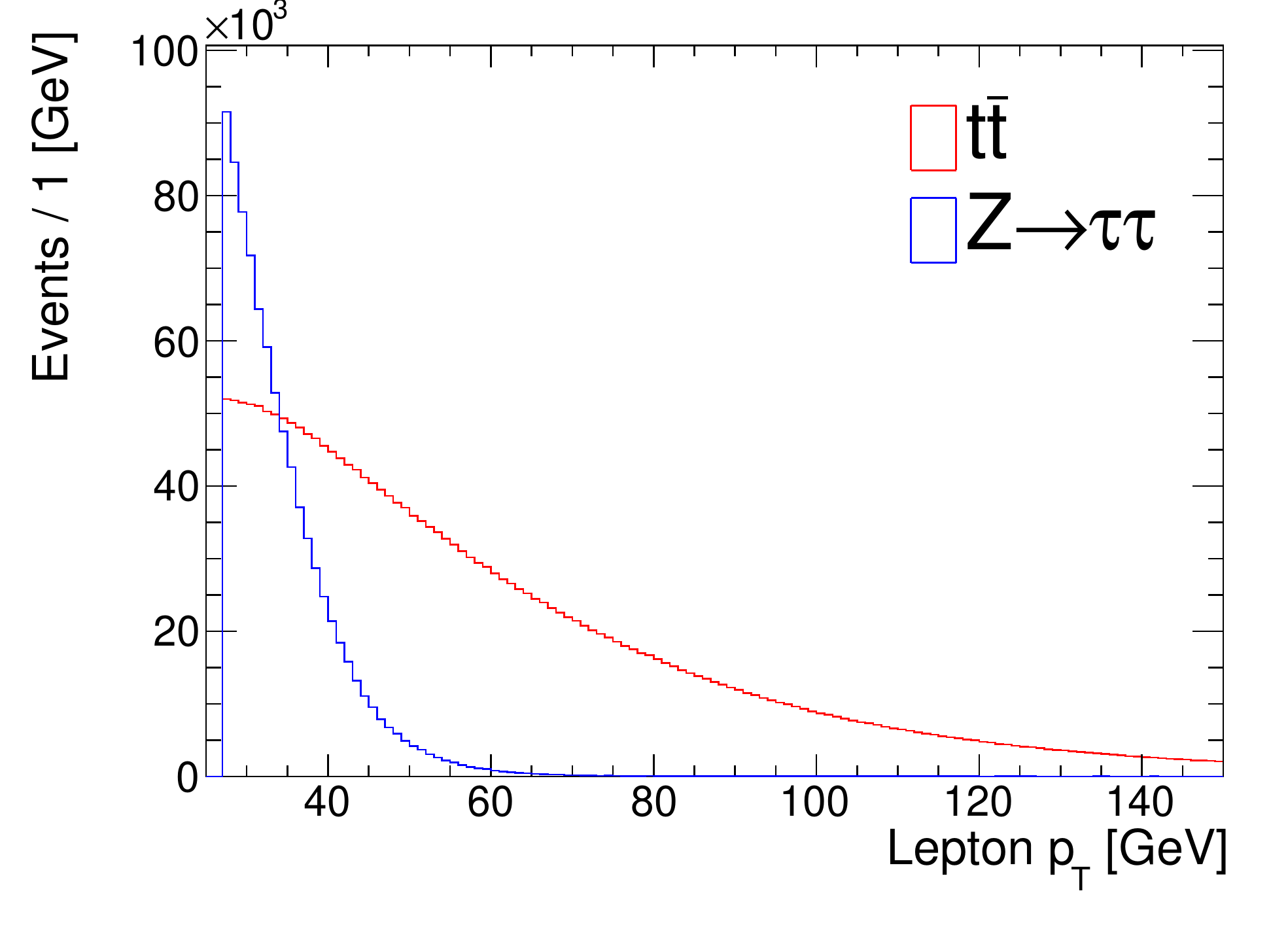}
\includegraphics[width=0.4\textwidth]{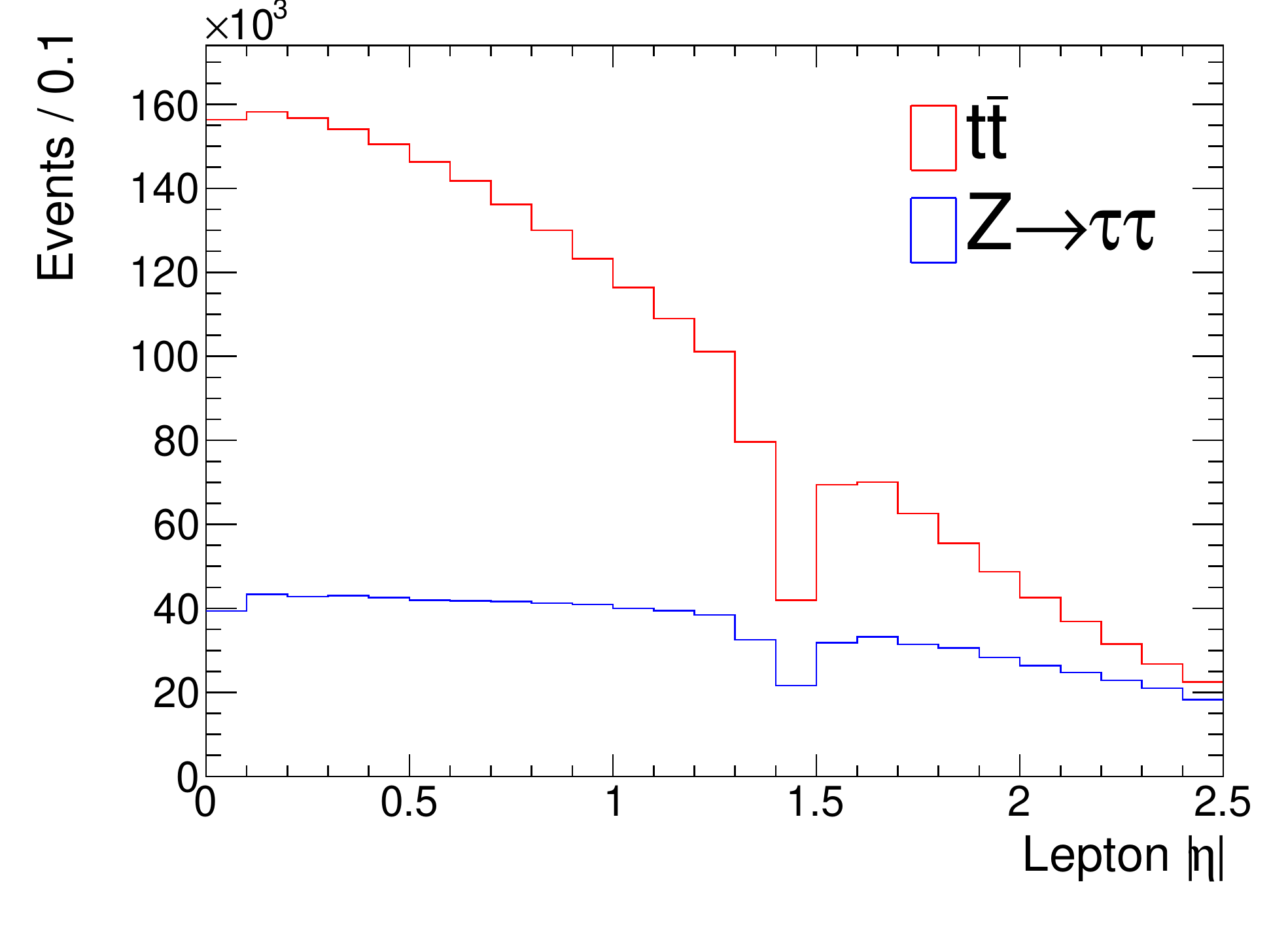}
\includegraphics[width=0.4\textwidth]{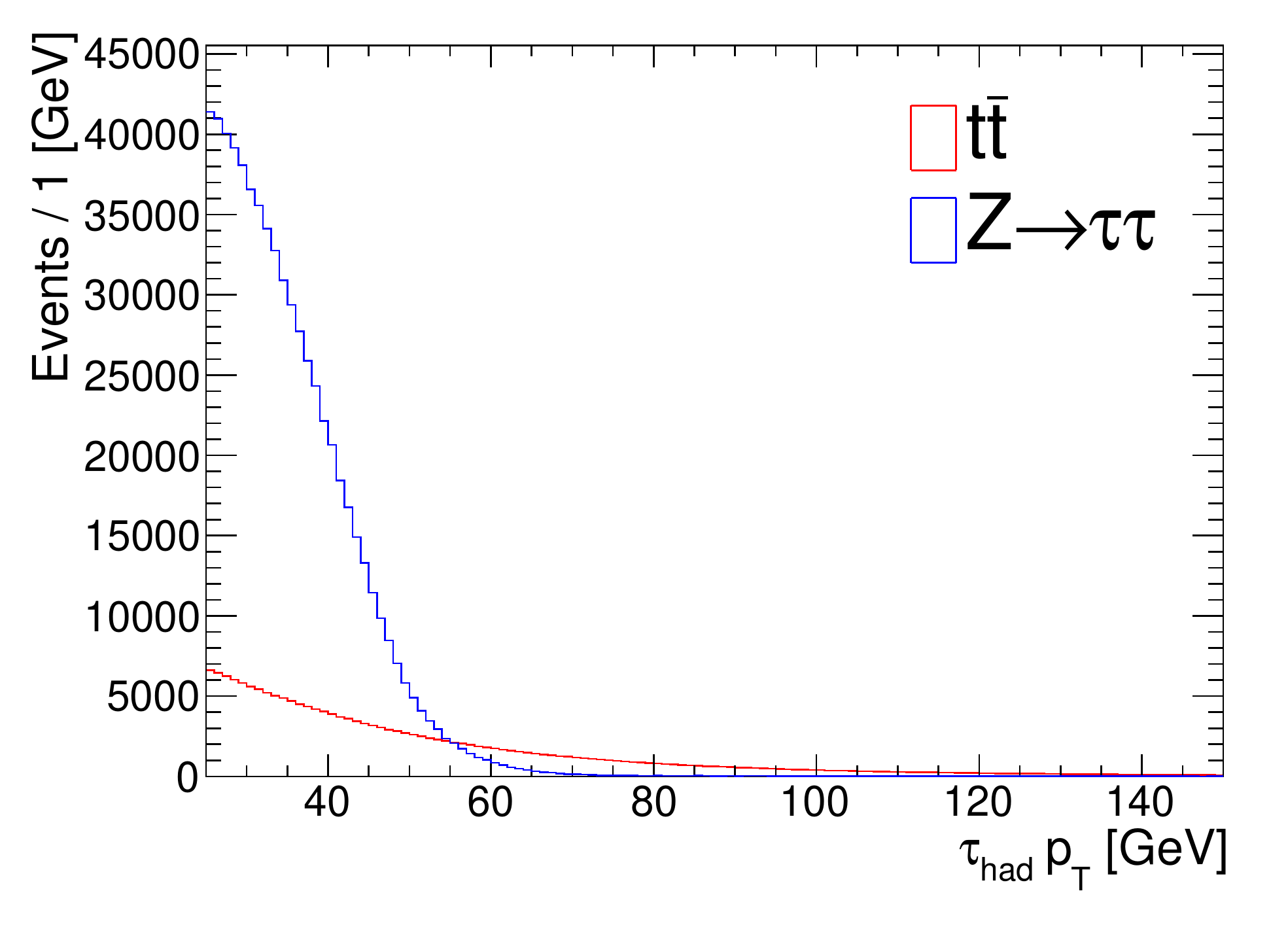}
\includegraphics[width=0.4\textwidth]{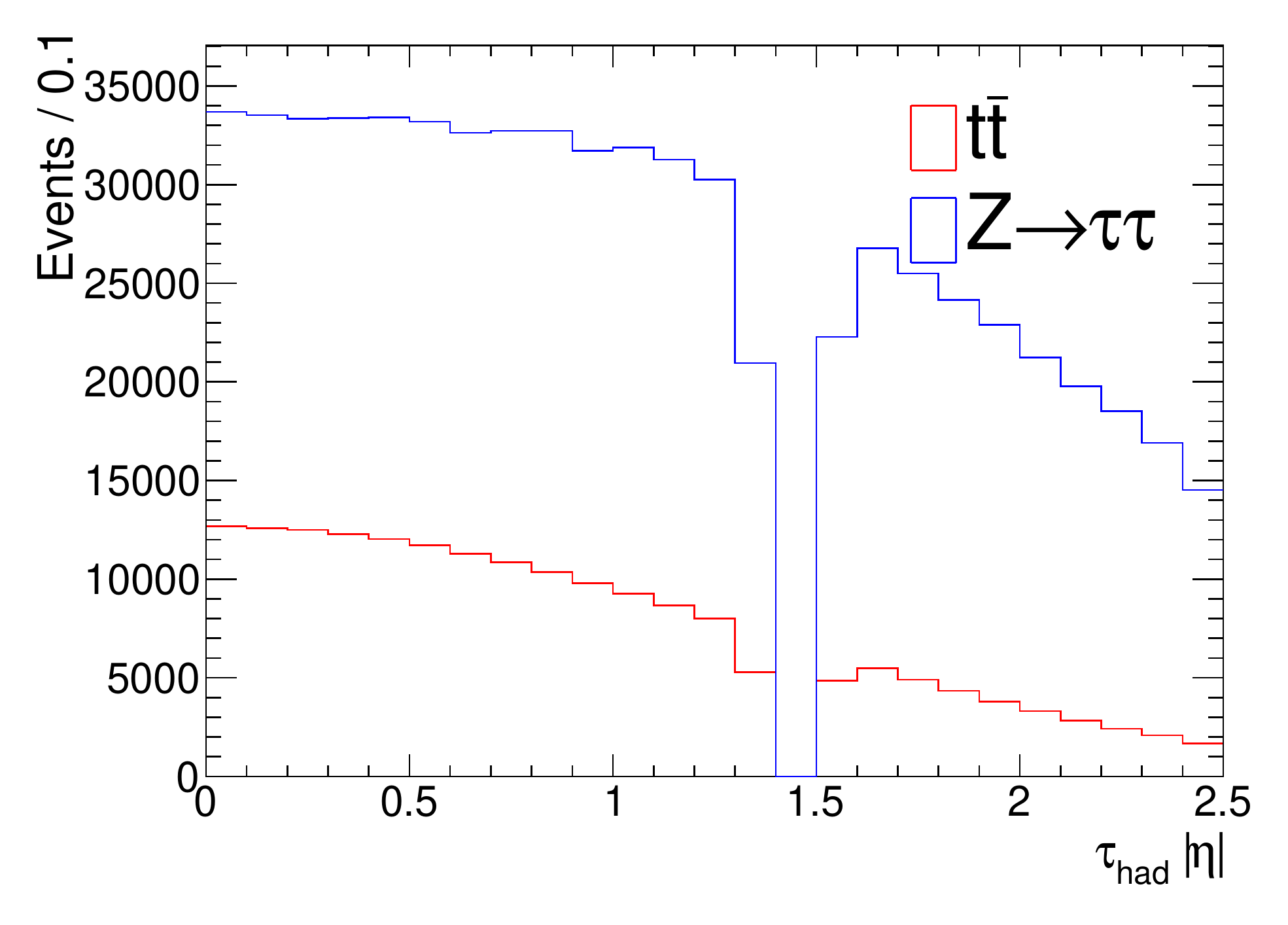}
\caption{
The distributions of $p_T$ and \modeta\ of  $e$, $\mu$, and \tauhad\ in the selected \ttWW\ and \Ztautau\ candidate event
samples.
The left column shows $p_T$ and the right column  shows \modeta.
The top row shows $e$ and $\mu$ candidates.
The bottom row shows \tauhad\ candidates.
}
\label{fig:ttbar-Ztt-pt-and-eta}
\end{figure*}

Figure~\ref{fig:ttbar-Ztt-pt-and-eta}
shows  the distributions of $p_T$ and \modeta\ of  $e$, $\mu$, and \tauhad, in the selected \ttWW\ (red) and \Ztautau\ (blue) candidate event samples.
It can be seen that the $p_T$ distributions are considerably softer for the \Ztautau\ candidate event samples than those for the selected \ttWW\ candidate event samples.

\begin{figure}[hbtp]
\centering
\includegraphics[width=0.4\textwidth]{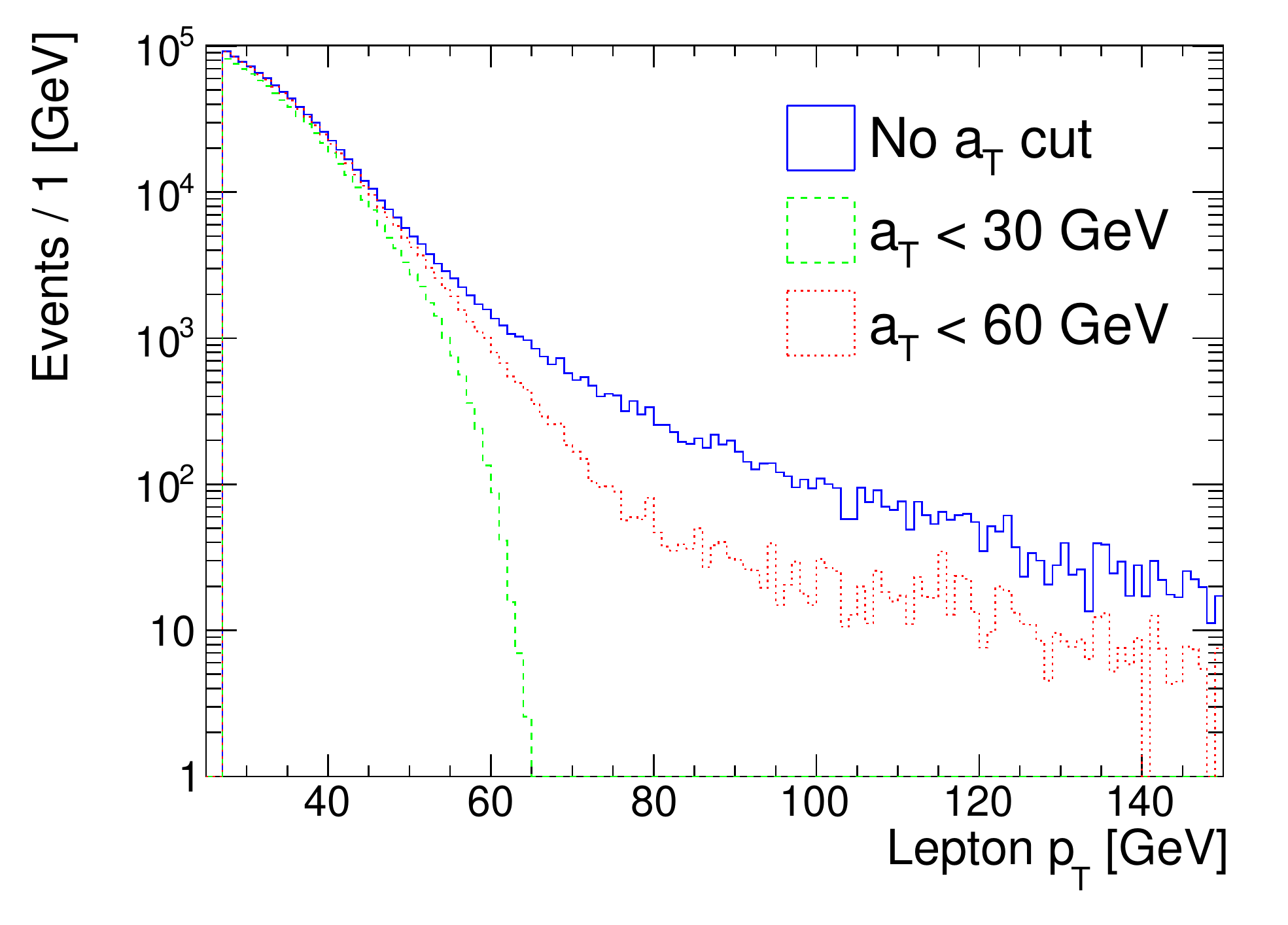}
\caption{Distributions of lepton $p_T$ in the selected \Ztautau\ candidate event samples.
Three different cuts on \at\ are applied.}
\label{fig:aT_cut}
\end{figure}

As can be seen from the lower plots in Figures~\ref{fig:Ztautau_selection}, a cut of $\at < 30$~GeV would be desirable to improve the suppression of backgrounds from \tt\ and EW diboson events.
However, a cut on \at\ suppresses also \Ztautau\ events that contain initial state parton radiation.
Initial state radiation broadens the distributions of lepton candidate $p_T$ in \Ztautau\ events.
A hard cut on \at\ would therefore suppress the high-$p_T$ regions in the distributions of lepton $p_T$ in the selected \Ztautau\ event samples, as is illustrated in Figure~\ref{fig:aT_cut}.
The cut  $\at < 60$~GeV is chosen to reject background events from \tt\ and diboson events, without unduly biasing the lepton
$p_T$ distributions and further accentuating the differences between the lepton $p_T$ distributions seen in
Figure~\ref{fig:ttbar-Ztt-pt-and-eta}.

\subsection{Size and composition of the selected event samples}    
\label{sec:MCsamples}

The expected numbers of selected events and the composition of the four selected event samples for \intL~=~140~\invfb\ at
$\sqrt{s}$~=~13~TeV are given in Table~\ref{tab:samples}.
\begin{table*}[hbtp]
\centering
\begin{tabular}{|c|cccc|}
\hline
Process & \multicolumn{4}{|c|}{Selected event sample} \\
\cline{2-5}
& \ttWWltau & \ttWWemu & \Zltau & \Zemu \\
\hline
\ttWW\ true & \bf178270 & \bf1092395 & 2034 & 6517 \\
\ttWW\ fake & 4761 & 0 & 20 & 156 \\
\tWb\ true & \bf3236 & \bf18209 & 0 & 725 \\
\tWb\ fake & 74 & 0 & 1 & 8 \\
Other single top & 1731 & 0 & 2 & 42 \\
\Ztautau\ true & 3 & 256 & \bf657280 & \bf101074 \\
\Zemutau\ fake & 0 & 0 & 1 & 14 \\
$W$+jets & 61 & 0 & 1226 & 0 \\
$WW$ true & 27 & 166 & 1555 & 6580 \\
Other diboson & 5 & 25 & 337 & 773 \\
\hline
Total background & 6663 & 447 & 5176 & 14815 \\
\hline
\end{tabular}
\caption{The expected numbers of events in the four selected event samples for \intL~=~140~\invfb\ at $\sqrt{s}$~=~13~TeV.
The numbers are broken down by physics production process.
The numbers for \tt, single top and $Z$ boson production are further broken down into the two categories ``true'', in which the two lepton
candidates are correctly identified, and ``fake'', in which at least one of the lepton candidates is incorrectly identified.
Small numbers of events from \ttV\ processes that contain two correctly identified leptonic $W$ boson decays are included in the \tt\ ``true' category.
Otherwise, the events from \ttV\ processes are included in the \tt\ ``fake'' category.
The numbers given in {\bf bold} type represent the signal in the four selected event samples.
The numbers given under ``Total background'' include the ``fake'' categories described above.
}
\label{tab:samples}
\end{table*}

As a result of this study we estimate that the fractional statistical uncertainty on \RWZ\ for a single LHC experiment for an integrated
luminosity of \intL~=~140~\invfb\ would be around 0.5\%.

The selection requirements for \ttWWltau\ and \ttWWemu\ can be seen to be very effective at removing non-\tt\ sources of background,
e.g., $W$+jet.
The most significant source of  non-\tt\ events originates from the EW production of ``single top'' events in the associated
production \tWb\ channel.
Since these events contain two genuine leptonically  decaying $W$ bosons they can effectively be considered as contributing to the signal sample, and not as background.
They are listed as ``\tWb\ true'' in Table~\ref{tab:samples}.
Single top processes that do not contain a pair of $W$ bosons decaying to produce two correctly identified leptons are classified as background.
Events which contain the associated production of either a $W$ or $Z$ boson with a \tt\ pair (\ttV) can be considered signal if they contain at least two $W$ bosons which decay to correctly identified leptons.
Since the fractions of these events are small, $<0.1\%$, they are included in the \tt\ categories in Table~\ref{tab:samples}.

The most significant source of background for \ttWWltau\ originates from genuine \tt\ events in which  the \tauhad\ candidate originates from a misidentified hadronic jet.
The residual background from this source corresponds
to about 2.5\% of the selected sample of candidate \ttWWltau\ events. 
The \ttWWemu\ candidate event sample will be selected with entirely
negligible levels of background.

The most significant source of background for \Zltau\ originates from
QCD {\it multijet}
(MJ) events (that is, events that do not contain any prompt leptons from $W$ or $Z$ boson decay).
This background is estimated to be at the level of around 5\%, as described in the Appendix and~\cite{bib:cms-taumj}.
In comparison, the MC-estimated backgrounds for \Zltau\ from  \tt\ (0.3\%) and $W$+jet (0.2\%) are small.
The most significant sources of background for \Zemu\  originate from \tt\ (5.8\%) and diboson (6.3\%) production.

\subsection{Sensitivity of $\boldsymbol{\RWZ}$ to $\boldsymbol{\RW}$}    
\label{sec:MCsensitivity}

\begin{figure}[hbtp]
\centering
\includegraphics[width=0.4\textwidth]{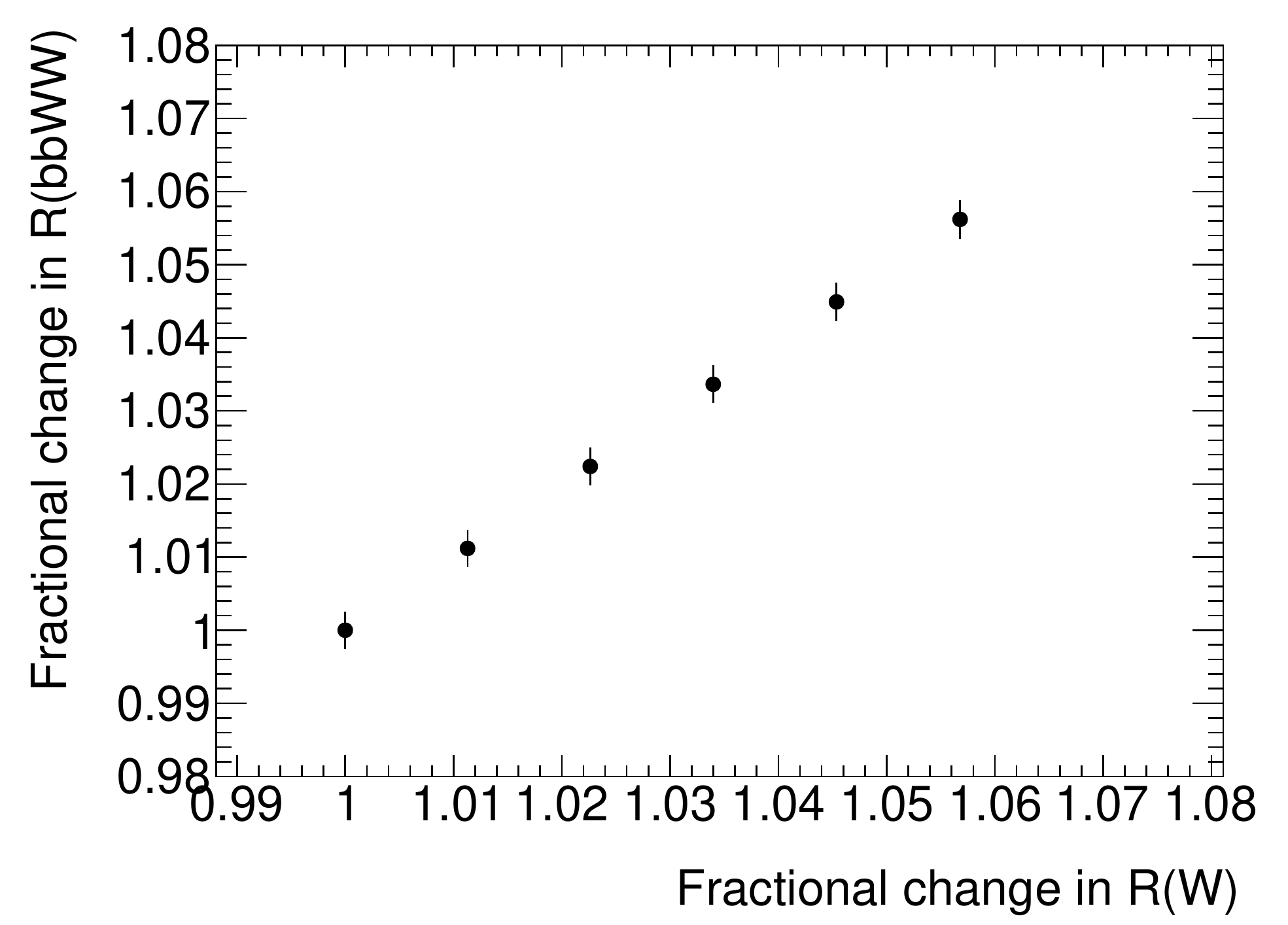}
\includegraphics[width=0.4\textwidth]{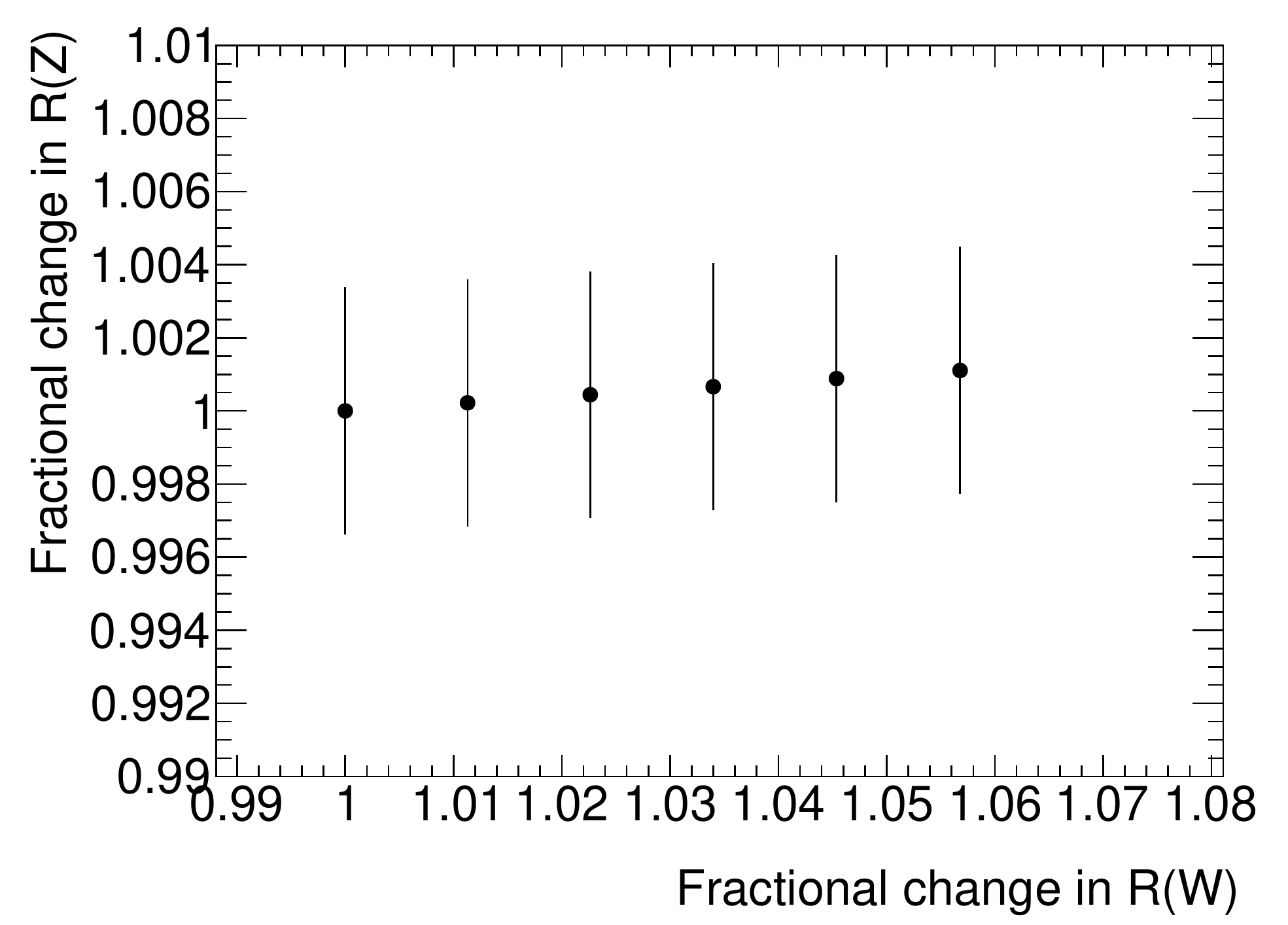}
\includegraphics[width=0.4\textwidth]{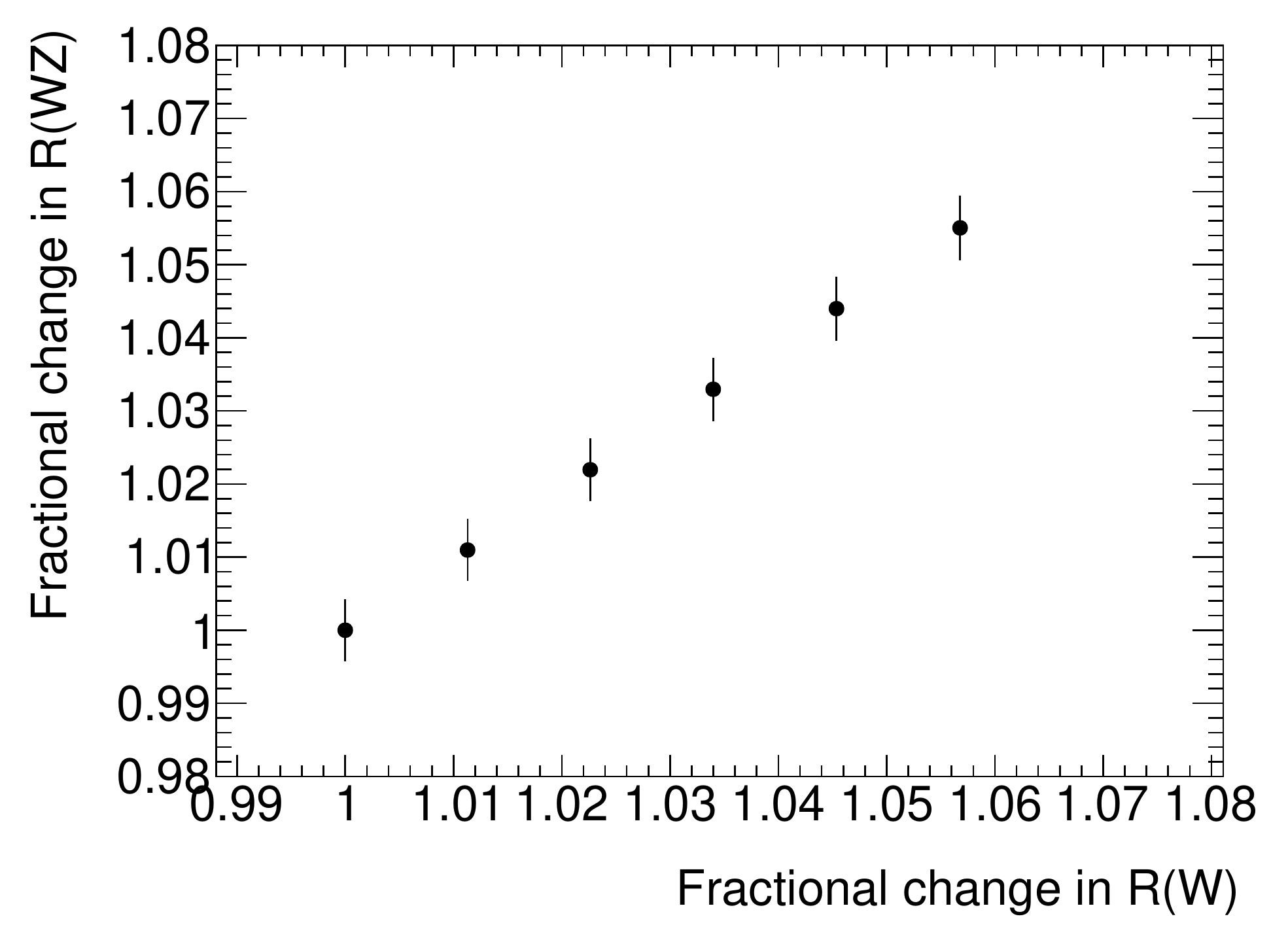}
\caption{The fractional change in  \RbbWW\ (upper), \RZ\ (middle), and \RWZ\ (lower),
  as a function of the fractional change in \RW, obtained by reweighting events at generator level as
  described in the text. The displayed error bars illustrate the expected experimental
  statistical uncertainty on the relevant quantity, and are completely correlated point to point.
  MC statistical uncertainties on the point-to-point variation with \RW\ are negligible.
}
\label{fig:MCsensitivity}
\end{figure}

For definiteness we make the assumption in our study that the effective \Wtaunu\ coupling relevant for on-mass-shell $W$ boson decays could be
modified by some BSM effect, whilst all other $W$ boson couplings are maintained at
their SM-predicted values.
Under this assumption, if the branching ratio $\mathcal{B}(\Wtaunu)$ is multiplied by a factor $X$ relative to its SM-predicted
value $\mathcal{B}(\Wtaunu)_{\mathrm{SM}}$,
\begin{equation}
 \mathcal{B}(\Wtaunu) = X.\mathcal{B}(\Wtaunu)_{\mathrm{SM}}
\label{eqn:x}
\end{equation}
then all
other $W$ boson branching fractions will be modified by a factor
\begin{equation}
  F = \frac{1-X.\mathcal{B}(\Wtaunu)_{\mathrm{SM}}}{1-\mathcal{B}(\Wtaunu)_{\mathrm{SM}}}.
\label{eqn:f}
\end{equation}
In our MC study we perform a ``calibration'' of the double-ratio method by reweighting every simulated event containing one or more
$W$ boson decays by a factor $X^nF^m$, where $n$ is the number of generator-level \Wtaunu\ decays and $m$ is the number of other $W$ boson decays.
This calibration procedure properly takes into account all events in the calculation of \RbbWW\ that contain pairs of leptonically decaying $W$ bosons with
correctly identified decay products in the ``numerator'' \ttWWltau\ and ``denominator'' \ttWWemu\ samples.
This includes, for example, the presence of events containing the cascade decay \Wtaunul, whose presence in the ``denominator'' \ttWWemu\ sample
slightly decreases the correlation between \RbbWW\ and \RW.
The value of \RZ\ is designed to be independent of any change in \RW. However, the backgrounds from \ttWW\ and $WW$\ in the \Zltau\ and \Zemu\
events contain decays of $W$ bosons, which cause the expected background levels to alter with \RW. A correction to the \RZ\ calculation from this effect is included.
The result of this calibration is shown in 
Figure~\ref{fig:MCsensitivity}, which shows the fractional change in   \RbbWW,\ \RZ,\ and \RWZ\ as a function of the fractional
change in \RW.
The absolute value of the ratio of \RW\ and \RWZ\ will depend on the precise experimental details specific to a given experiment and
would have to be evaluated using fully simulated MC events for the specific identification criteria and event selection cuts
employed in the analysis of the experimental data.

\subsection{Evaluation of systematic uncertainties}    
\label{sec:MCsyst}

Considering each source of systematic uncertainty described in sections~\ref{sec:MCdetector} and~\ref{sec:MCgenerators}, and summarised in
Tables~\ref{tab:systematics} and~\ref{tab:pheno-systematics}, we evaluate the resulting changes in the
ratios  \RbbWW, \RZ, \RWZ,  and \RW.

\begin{figure}[hbtp]
\centering
\subfigure[\RbbWW]{\includegraphics[width=0.5\textwidth]{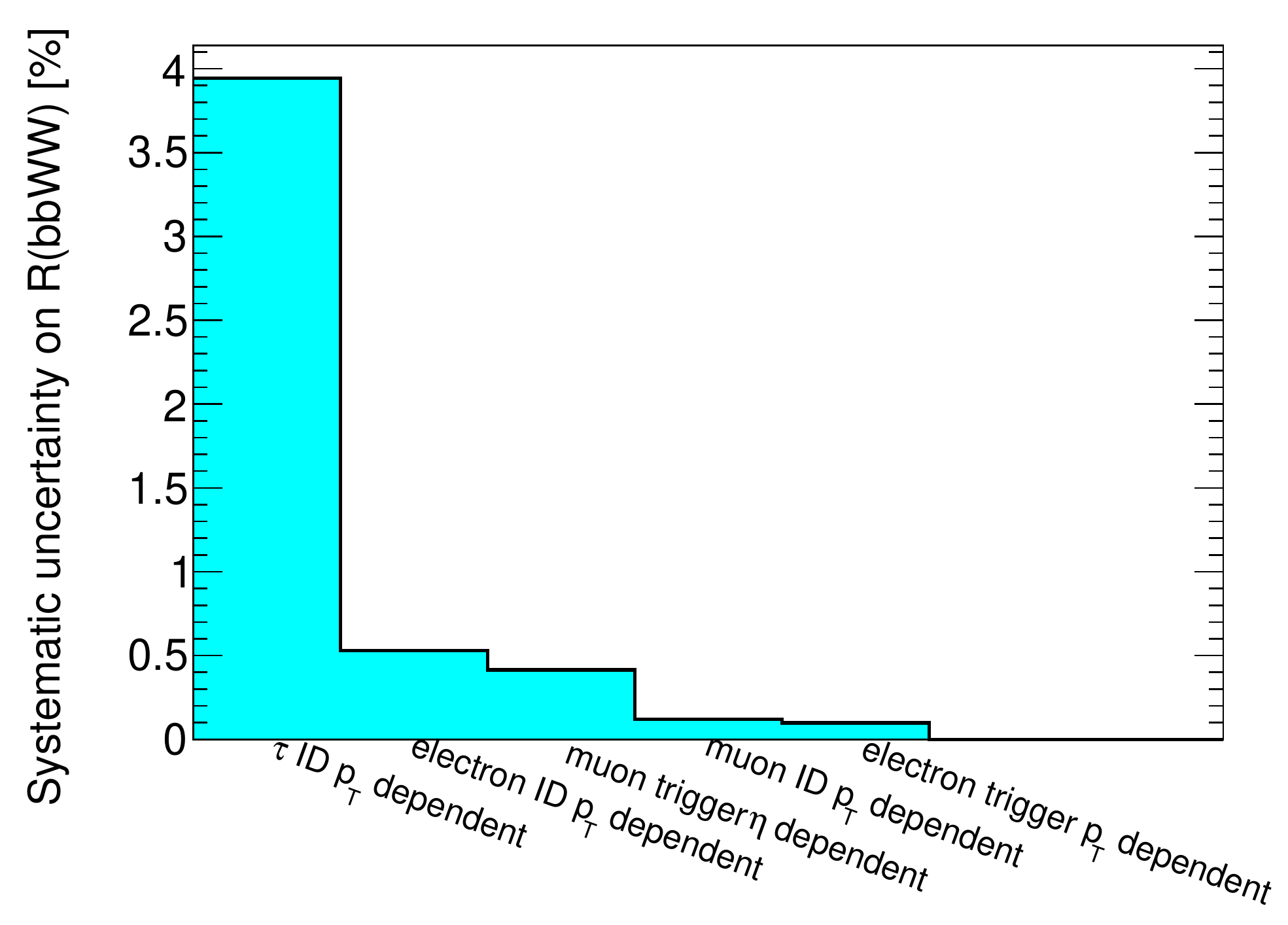}}
\subfigure[\RZ]{\includegraphics[width=0.5\textwidth]{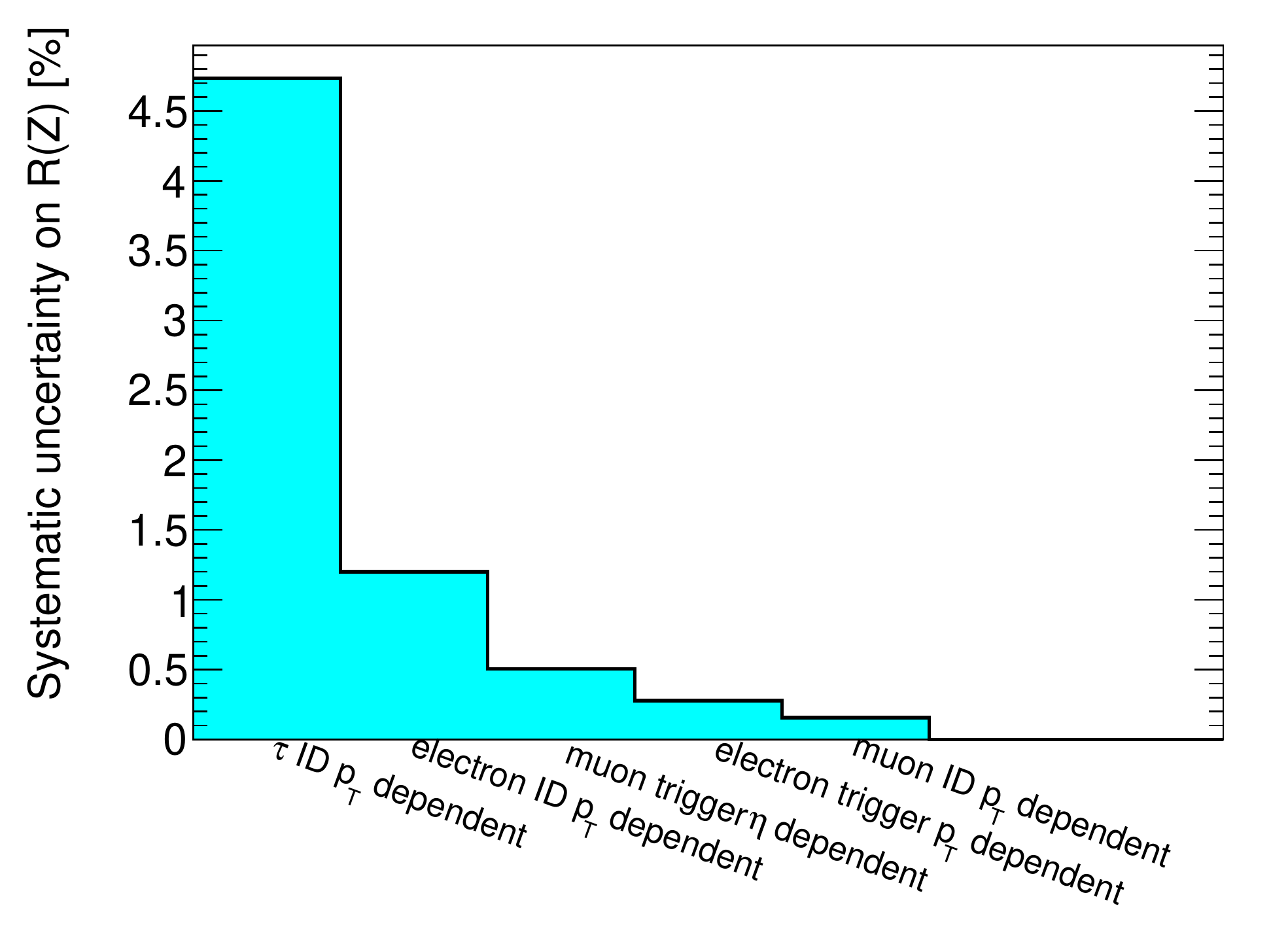}}
\subfigure[\RWZ]{\includegraphics[width=0.5\textwidth]{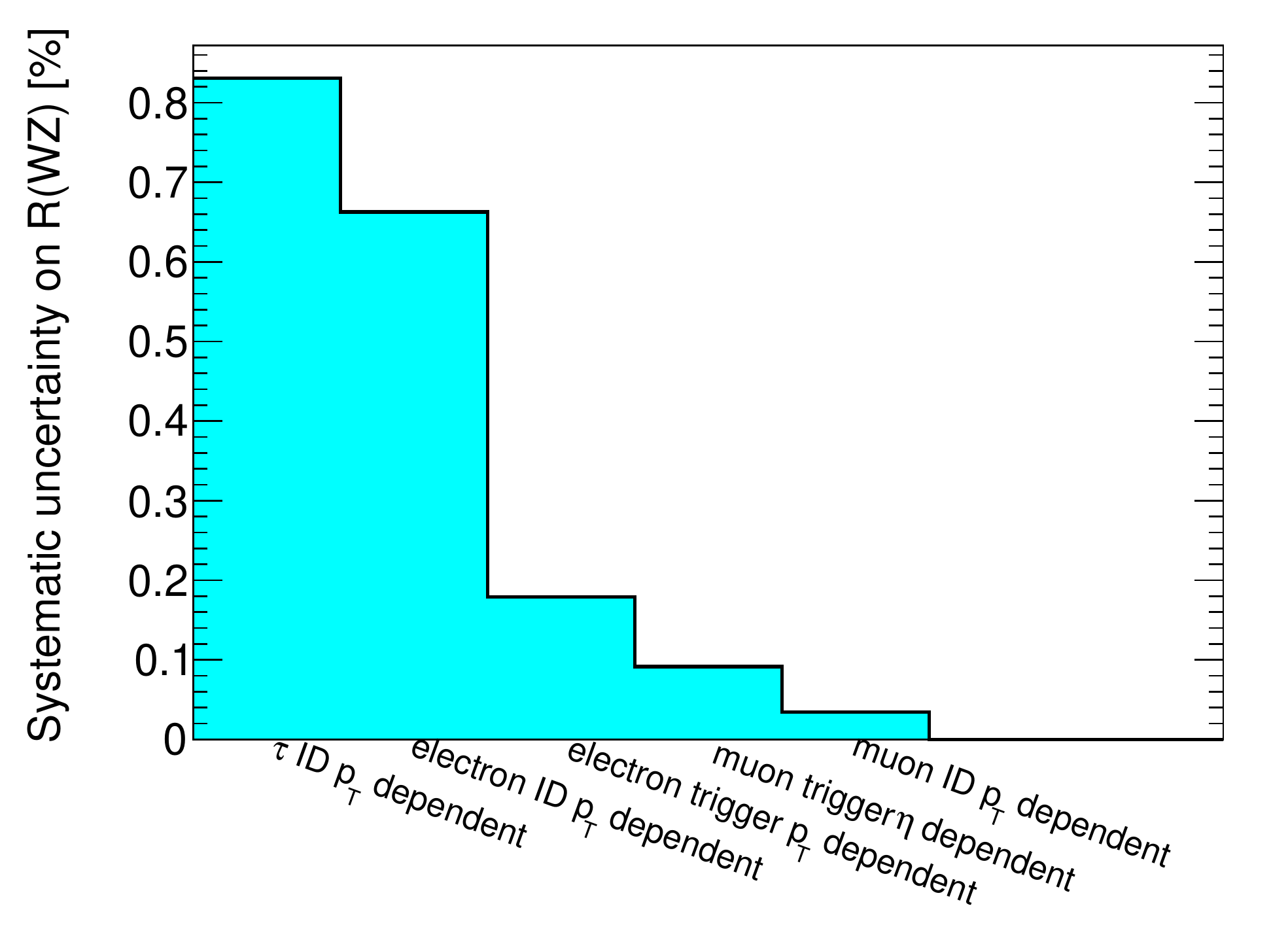}}
\caption{Summary of the most significant changes in the ratios
  (a)~\RbbWW, (b)~\RZ, and (c)~\RWZ\ arising from sources of $p_T$/$\eta$-dependent systematic uncertainty.
  From left to right in each sub-figure the sources are ordered in descending magnitude of the resulting systematic uncertainty.
}
\label{fig:syst}
\end{figure}

The $p_T$/$\eta$-independent and $p_T$/$\eta$-dependent systematic variations on the reconstruction of leptons and jets, as summarised in
Table~\ref{tab:systematics}, are considered separately.
The resultant changes in the ratios are given in Table~\ref{tab:results}.
It can be seen that the  $p_T$/$\eta$-independent systematic uncertainties on \RbbWW\ and \RZ\ are large,
but almost exactly equal.
Therefore, $p_T$/$\eta$-independent systematic uncertainties on the reconstruction of leptons and jets almost perfectly cancel in the
double ratio \RWZ.
When considering $p_T$/$\eta$-dependent systematic uncertainties the cancellation is no longer perfect.
The most significant sources of $p_T$/$\eta$-dependent systematic uncertainties are illustrated in Figure~\ref{fig:syst}.
The 10\% uncertainty on the 2.5\% background from hadronic jets misidentified as \tauhad\ candidates in the \ttWWltau\ sample leads to an uncertainty of 0.25\% on
\RbbWW\ and thus also on \RW.
The 5\% uncertainty on the 5\% background from MJ events in the \Zltau\ sample leads to an uncertainty of 0.25\% on
\RZ\ and thus also on \RW.
Since the above two backgrounds both result from hadronic jets misidentified as \tauhad\ candidates it is conceivable that the
resultant systematic uncertainties on \RbbWW\ and \RZ\ could be correlated and thus partially cancel in the calculation of \RW.
A realistic estimate of the degree of correlation will depend on experimental details beyond the scope of the current study and we
have not taken into account any potential reduction in the systematic uncertainty on \RW.
\begin{table*}[hbtp]
\centering
\begin{tabular}{|c|c|c|c|c|c|c|}
\hline
Source of &  \multicolumn{6}{|c|}{Systematic uncertainty on measured ratios (\%)} \\
\cline{2-7}
systematic & \multicolumn{3}{|c|}{$p_T$/$\eta$-independent variation} & \multicolumn{3}{|c|}{$p_T$/$\eta$-dependent variation} \\
\cline{2-7}
uncertainty & \RbbWW\ \ \ & \ \ \ \RZ\ \ \ & \ \ \ \RWZ\ \ \ & \ \ \ \RbbWW\ \ \ & \ \ \ \RZ\ \ \ & \ \ \ \RWZ\ \ \ \\
\hline
Tau ID & $5.0$ & $5.0$ & $<0.1$ & $3.9$ & $4.7$ & $0.8$ \\
Electron ID & $1.1$ & $1.3$ & $0.2$ & $0.5$ & $1.2$ & $0.7$ \\
Electron trigger & $0.3$ & $0.2$ & $0.1$ & $0.1$ & $0.3$ & $0.2$ \\
Muon trigger & $0.4$ & $0.5$ & $0.1$ & $0.4$ & $0.5$ & $0.1$ \\
Muon ID & $1.0$ & $0.7$ & $0.3$ & $0.1$ & $0.2$ & $<0.1$ \\
b-jet ID & $<0.1$ & $<0.1$ & $<0.1$ & - & - & -  \\
Light jet mis-ID & $0.1$ & $0.1$ & $<0.1$ & - & - & -  \\
Tau $p_T$ scale & $<0.1$ & $<0.1$ & $<0.1$ & $<0.1$ & $<0.1$ & $<0.1$ \\
Electron $p_T$ scale & $<0.1$ & $<0.1$ & $<0.1$ & $<0.1$ & $<0.1$ & $<0.1$ \\
Muon $p_T$ scale & $<0.1$ & $<0.1$ & $<0.1$ & $<0.1$ & $<0.1$ & $<0.1$ \\
Jet energy scale & $<0.1$ & $<0.1$ & $<0.1$ & - & - & - \\
Electron $p_T$ resolution & $<0.1$ & $<0.1$ & $<0.1$ & $<0.1$ & $<0.1$ & $<0.1$ \\
Muon $p_T$ resolution & $<0.1$ & $<0.1$ & $<0.1$ & $<0.1$ & $<0.1$ & $<0.1$ \\
Fake \tauhad\ background & & & & & & \\ (\ttWWltau\ sample) & 0.25 & 0 & 0.25 & - & - & -  \\
MJ background & & & & & & \\ (\Zltau\ sample) & 0 & 0.25 & 0.25 & - & - & -  \\
\hline
\end{tabular}
\caption{Changes in the
ratios  \RbbWW, \RZ, and \RWZ\ resulting from the sources of experimental systematic uncertainty described in Section~\protect\ref{sec:MCdetector} and summarised in
Table~\protect\ref{tab:systematics}.
The $p_T$/$\eta$-independent and $p_T$/$\eta$-dependent systematic variations are considered separately.
}
\label{tab:results}
\end{table*}

The sources of phenomenological systematic uncertainty considered in this study are summarised in Table~\ref{tab:pheno-systematics} and the
resultant changes in the ratios \RbbWW, \RZ, and \RWZ\ are given in Table~\ref{tab:pheno-results}.
The 3\% uncertainty on the backgrounds from  \tt\ (5.8\%) and diboson (6.3\%) production in the \Zemu\ sample leads to uncertainties
of 0.2\% and 0.2\%, respectively, on
\RZ\ and thus also on \RW.
The uncertainty resulting from the \Zpt\ reweighting procedure is $<$0.1\% on \RZ, and thus also on \RW.

The alternative \tt\ sample with modified QCD factorisation and
renormalisation scales leads to an uncertainty of 0.3\% on \RbbWW\ and thus also on \RW~\cite{bib:binomial}.
The \mtt\ reweighting leads to an uncertainty of $0.2\%$ on \newline\RbbWW\ and thus also on \RW.
\begin{table*}[tphb]
\centering
\begin{tabular}{|c|c|c|c|}
\hline
Source of systematic uncertainty &  \multicolumn{3}{|c|}{Systematic uncertainty on measured ratios (\%)} \\
\cline{2-4}
 &  \RbbWW & \ \ \ \RZ\ \ \ & \ \ \RWZ\ \ \\
\hline
\tt\ background in  \Zemu\ sample & 0 & 0.2 & 0.2 \\
Diboson background in  \Zemu\ sample & 0 & 0.2 & 0.2 \\
\Zpt\ reweighting & 0 & $<0.1$ & $<0.1$ \\
\mtt\ reweighting & 0.2 & 0 & 0.2 \\
\tt\ modelling & 0.3 & 0 & 0.3 \\
\hline
\end{tabular}
\caption{Changes in the
ratios  \RbbWW, \RZ,  and \RWZ\ resulting from the sources of phenomenological  systematic uncertainty described in Section~\protect\ref{sec:MCgenerators} and summarised in
Table~\protect\ref{tab:pheno-systematics}.
}
\label{tab:pheno-results}
\end{table*}

We add in quadrature the changes in the double ratio \RWZ\ arising from the $p_T$/$\eta$-independent and $p_T$/$\eta$-dependent
systematic variations on the reconstruction of leptons and jets, together with the other considered systematic uncertainties discussed above.
The total resulting systematic uncertainty on \RWZ\ is \RWZPrecisionSyst.

\subsection{Considerations for future measurements}    
\label{sec:MClimitations}

Measurements of \RW\ on the data from ATLAS and CMS using the double ratio technique proposed here will, clearly, require the
use of fully simulated MC events and will employ the sophisticated procedures developed by the individual experiments to evaluate backgrounds and
experimental systematic uncertainties.
Our study is based on particle-level MC events and a simple parameterised simulation of detector performance.
Nevertheless, we believe our
demonstration that the dominant experimental systematic uncertainties cancel in the double ratio, as well as our estimates of the major residual
systematic uncertainties, to be broadly realistic.
We have based our simulation of lepton, jet and \etmiss\ reconstruction on measurements of efficiencies, backgrounds and systematic
uncertainties published by ATLAS and CMS~\cite{bib:atlas-muonID}--\cite{bib:pile-up}, having chosen identification algorithms whose
performance is suited to the needs of our analysis.
The above cited performance papers are for the most part based on around a quarter of the full run~2 data set of \intL~=~140~\invfb\ at
$\sqrt{s}$~=~13~TeV that is now available. 
It is, therefore, to be expected that the full run~2 data set will allow systematic uncertainties on
lepton and jet identification efficiencies to be reduced by about a factor of two compared to the values we have assumed in our study.
Of particular relevance to our study, it is to be hoped that the high statistics provided by the full run~2 data set will
allow  the $p_T$ dependence of the identification efficiency for  \tauhad\ candidates to be studied over the range $25<p_T<100$~GeV,  without the need to use \ttWWltau\ events and
implicitly assume that $\mathcal{B}(\Wtaunu)$ takes its SM value.
This could be achieved, e.g., by the selection of a dedicated \Zltau\ event sample in which there is a high transverse momentum
initial state radiation.

The residual non-cancellation in \RWZ\ of the $p_T$/$\eta$-dependent  \tauhad\ identification uncertainty, as shown in
Table~\ref{tab:results}, arises primarily from the differences in the \tauhad~$p_T$ distributions between the \ttWWltau\ and \Zltau\ samples, as shown in Figure~\ref{fig:ttbar-Ztt-pt-and-eta}.
We have considered possible methods to reduce the impact of the different $\tau$~$p_T$ distributions.
One simple approach would be to introduce an additional cut on \tauhad~$p_T < 65$~GeV; this removes the high $p_T$ tail in the \ttWWltau\ sample, with
negligible changes to the \Zltau\ sample.
This has the effect of improving the cancellation in \RWZ\ of the $p_T$-dependent \tauhad\ identification uncertainty between the
\ttWWltau\ and \Zltau\ samples;
the residual systematic uncertainty reduces from 0.8\%, as shown in
Table~\ref{tab:results}, to 0.2\%.
More sophisticated potential methods to mitigate the effects of the $p_T$-dependent  \tauhad\ identification uncertainty might be
(a)~to reweight the tau $p_T$ distribution in the \ttWWltau\ sample to resemble that in the \Zltau\ sample, or
(b)~to perform measurements of \RWZ\ in bins of  \tauhad~$p_T$ and subsequently combine the measurements, taking into account the
bin-to-bin correlations in the systematic uncertainties.
However, any such methods might run the risk of increasing the sensitivity of \RWZ\ to phenomenological
uncertainties.
A careful investigation of such issues will be needed in order to optimise the overall uncertainty arising from experimental and phenomenological systematic
uncertainties.
This will require study of the large number of fully simulated MC samples produced by the experiments, corresponding to different models of
\tt\ and $Z$ boson production and decay, and is beyond the scope of this paper.

Of course, if the measured value of \RWZ\ is found to disagree with that expected in the SM then further studies will be required to
ascertain the nature of BSM physics that is responsible.
For example, the decay $t\rightarrow bH^+$, where $H^+$ is a charged higgs boson, followed by $H^+\rightarrow \tau^+\nu$ would modify the effective \ttotaunu\ and
\ttolnu\ branching ratios in a similar fashion to that discussed in the context of equations~\ref{eqn:x} and~\ref{eqn:f} above.
Similarly, top decays via a neutral higgs boson $t\rightarrow qH$, $H\rightarrow \tau\tau$ will also increase the number of tau
leptons in \tt\ events relative to the SM-expected value.
Existing experimental searches for charged~\cite{bib:atlas-t-charged-higgs,bib:cms-t-charged-higgs} and neutral~\cite{bib:atlas-t-neutral-higgs} higgs bosons in top quark events suffer from the large systematic uncertainties associated with
\tauhad\ identification and will benefit from the double ratio method we propose in this paper to reduce experimental systematic uncertainties.

The large numbers of events from the EW production of diboson events ($WW$ and $WZ$) at the LHC provide alternative samples with which to make this novel measurement.
Controlling systematic uncertainties on \RW\ 
in \WWlnutaunu\ events will be extremely challenging; we shall need extraordinarily good background rejection against fake $\tau$ candidates from 
misidentified jets in \Wlnu + jet events.
One may also consider  \ZWlltaunu\  as an alternative sample
with which to perform this measurement.
This channel provides 
lower statistics because of the lower cross section times branching fraction.
However, one can veto on $Z$+jet backgrounds by removing events in which the \ellell\ momentum is back to back with the
$\tau$ candidate direction.
One can calibrate the residual backgrounds by looking at the back-to-back events.
If the experimental measurements proposed here observe a clear violation of LU then having  three channels \ttWWltau, \WWlnutaunu,
and \ZWlltaunu\ could increase the significance of the observation, and could help elucidate the underlying BSM origin of the effect.

In addition to the \ttWWltau\ and \Zltau\ final states considered here, it may be possible to measure \RW\ using a similarly
motivated ratio of \ttWWltaulept\ and \Ztaulept\ final states, where \taulept\ corresponds to a leptonic decay of the $\tau$,
that may be identified using, for example, criteria based on the non-zero lifetime of the $\tau$ lepton.
A combination of \RW\ measurements using \tauhad\ and \taulept\ signatures would benefit from the fact that the leading
experimental uncertainties, arising from \tauhad\ and \taulept\ identification, would be largely uncorrrelated.


 \section{Summary and conclusions}
 \label{sec:conclusions}
 
A  measurement of 
$\RW \equiv \mathcal{B}(\Wtaunu)/\mathcal{B}(\Wlnu)$ ($\ell=e\ \mathrm{or}\ \mu$)
represents a promising opportunity to
discover a violation of lepton universality.
We propose here a novel double-ratio method that will allow \RW\ to be measured using top quark pairs and \Ztautau\ events at the LHC.

We define \RbbWW\ in di-leptonic \tt\ events to be the ratio of the numbers of \ltau\ and \emu\ final states
(equation~\ref{eqn:RbbWW}).
\RbbWW\ is sensitive to the value of \RW, but also to systematic uncertainties on the reconstruction of $e$, $\mu$, and $\tau$ leptons.
Similarly, we define \RZ\ in  \Ztautau\ events to be the ratio of the numbers of \ltau\ and \emu\ final states
(equation~\ref{eqn:RZ}).
\RZ\ is similarly sensitive to systematic uncertainties on the reconstruction of $e$, $\mu$, and $\tau$ leptons, but is insensitive to the value of \RW.
The double ratio $\RWZ \equiv \RbbWW / \RZ$ cancels to first order sensitivity to systematic uncertainties on the reconstruction of
$e$, $\mu$, and $\tau$ leptons, thus improving very significantly  the precision to which \RW\ can be measured at a hadron collider

We have performed a study of the double ratio \RWZ\ using particle-level MC events and a parameterised simulation of detector performance.
We have based our simulation of lepton, jet and \etmiss\ reconstruction on measurements of efficiencies, backgrounds and systematic
uncertainties published by ATLAS and CMS~\cite{bib:atlas-muonID}--\cite{bib:pile-up}, having chosen identification algorithms whose
performance is suited to the needs of our analysis.
For a data set of \intL~=~140~\invfb\ at
$\sqrt{s}$~=~13~TeV we estimate a statistical uncertainty on \RW\ of 0.5\%.
Our study confirms the almost perfect cancellation in \RW\ of systematic uncertainties on the reconstruction efficiencies of $e$,
$\mu$, and $\tau$ leptons that are applied as constant factors.
We find that the most significant residual sources of uncertainty on \RW\ arise from systematic uncertainties on the $p_T$ and $\eta$
dependence of the reconstruction efficiencies of $e$, $\mu$, and $\tau$ leptons, which total around $1.0\%$.
We have evaluated also potential uncertainties arising from backgrounds to the selected event samples and from various
phenomenological sources.

Our studies indicate that a single experiment precision on the measurement of \RW\ of around \RWZPrecisionStatSyst\ is achievable with a data set of \intL~=~140~\invfb\ at
$\sqrt{s}$~=~13~TeV.
This would improve significantly upon the precision of the LEP2 measurements of \RW.
If the central value of the new measurements were
equal to the central value of the LEP2 measurements this would yield an observation of BSM physics at a significance level
of around 5$\sigma$.

\begin{acknowledgement}
Acknowledgements: Final year undergraduate (MPhys) project students working in Manchester with T.W.~have made some important contributions to the ideas
presented in this paper.
Jihyun Jeong (1993--2018) and Robin Upham made an early feasibility study for the measurement of \RbbWW\ using \tt\ events at the LHC.
Vilius Cepaitis and Ricardo W\"olker studied possible improvements to the selection of \Ztautau\ events at the LHC from which the
variable \mstar\ arose.
We are very grateful to our colleague Chris Parkes for his useful suggestions for the improvement of this paper.
\end{acknowledgement}


\section*{Appendix: Detailed Description of the Parameterised Detector Simulation}
\renewcommand\thesection{\Alph{section}}
\setcounter{section}{1}

We give here a detailed description of our choices of lepton and jet identification algorithms to simulate, the experimental systematic uncertainties we
consider, and the reasoning behind these choices.

\subsection{Simulation of muon candidates}
\label{sec:MCmuons}
The efficiencies and systematic uncertainties associated with the reconstruction and identification of high $p_T$, isolated muon candidates have been
presented by the ATLAS~\cite{bib:atlas-muonID} and CMS~\cite{bib:cms-muonID} collaborations. 
In both experiments the muon reconstruction efficiency is around 99\% and is independent of $p_T$ and \modeta, except for some
well-defined, poorly instrumented regions of both detectors that are usually excluded for precision measurements.
We choose to simulate the efficiency for muon identification according to that given for the ``Medium'' category described
in~\cite{bib:atlas-muonID}.
This gives an efficiency of around 96\% for muons with $p_T>20$~GeV.
Systematic uncertainties on the efficiency are around 0.1\% for muon $p_T$ around $m_Z/2$, increasing to around 0.5\% for
$p_T\approx 30$~GeV and $p_T\approx 100$~GeV.
We choose to simulate the ``Tight'' lepton isolation requirement given in~\cite{bib:atlas-muonID}, which is measured to have an
efficiency that is independent of $p_T$ and \modeta\ of around 96\% with a systematic uncertainty at the per mille level.
We assume the efficiency of the isolation criteria for sources of non-prompt muons to be 0.03,  based on the range of values given in~\cite{bib:atlas-muonID}.   
We consider a combined systematic uncertainty on the efficiency for muon  reconstruction, identification, and isolation.
We consider a  $p_T$/$\eta$-independent relative systematic uncertainty of 2\%. 
We generate a  $p_T$-dependent systematic uncertainty  by modifying the efficiency by a relative amount
that varies linearly between 0.5\% at $p_T = 30$~GeV and 0\% at $p_T = 80$~GeV and above. 

We simulate the efficiency of the single muon trigger to be  65\% (80\%) in the barrel (endcap) region, as has been measured for ATLAS~\cite{bib:atlas-trigger,bib:cms-trigger}.
We define the barrel region by $|\eta|<1.0$ and the endcap region by $|\eta|>1.0$.
For offline $p_T$ at least 1~GeV above the trigger threshold the efficiency is almost independent of $p_T$.
The trigger threshold is assumed to be similar to the electron trigger threshold of ATLAS during most of run~2 at $p_T = 26$~GeV~\cite{bib:atlas-triggerthreshold-electron}.
We consider a  $p_T$/$\eta$-independent relative systematic uncertainty of 1\%. 
We generate an $\eta$-dependent systematic uncertainty  by modifying the  relative efficiency by 1\% in the endcap region,
whilst the efficiency in the barrel region remains unchanged.

We simulate the resolution in muon $p_T$ by means of a Gaussian with a width of $2.30\pm 0.15$\% in the barrel region and  $2.90\pm 0.15$\% in the endcap
region, independent of $p_T$~\cite{bib:atlas-muonID}.
We consider a  $p_T$/$\eta$-independent relative systematic uncertainty of 0.2\% on the $p_T$ scale~\cite{bib:atlas-muonID}.
We generate $\eta$-dependent systematic uncertainties  by modifying the resolution and $p_T$ scale only in the endcap region,
whilst the values in the barrel region remain unchanged.

\subsection{Simulation of electron candidates}
\label{sec:MCelectrons}
The efficiencies and systematic uncertainties associated with the reconstruction and identification of high $p_T$, isolated electron candidates have been
presented by the ATLAS~\cite{bib:atlas-electronID} and CMS~\cite{bib:cms-electronID} collaborations. 
In both experiments the electron reconstruction efficiency is around 98\% and is fairly independent of $p_T$ and \modeta, except for some
well-defined, poorly instrumented regions of both detectors~\cite{bib:atlas-elec-eta-crack}.
A general feature in both ATLAS and CMS is that the efficiency of the commonly used electron identification algorithms tend to have
a larger dependence on $p_T$ than is the case for muon identification.
We choose to simulate approximately the efficiency for electron identification according to the ``Tight likelihood'' algorithm of~\cite{bib:atlas-electronID}.
The efficiency is around 75\% at $p_T=30$~GeV and rises to
around 90\% for $p_T>80$~GeV and above~\cite{bib:atlas-electronIDeff}.
The efficiency is determined with a systematic uncertainty of a few per mille for $p_T>30$~GeV.
We choose to simulate also for electrons the ``Tight'' lepton isolation requirement given in~\cite{bib:atlas-muonID}, with an
efficiency that is independent of $p_T$ and \modeta\ of around 96\%.
We consider a combined systematic uncertainty on the efficiency for electron  reconstruction, identification, and isolation.
We consider a  $p_T$/$\eta$-independent relative systematic uncertainty of 2\%. 
We generate a  $p_T$-dependent systematic uncertainty  by modifying the efficiency by a relative amount
that varies linearly between 2\% at $p_T = 30$~GeV and 0\% at $p_T = 80$~GeV and above. 

The efficiency of the single electron trigger has been measured in ATLAS to be around 90\%
for offline $p_T$ at least 1~GeV above the trigger threshold~\cite{bib:atlas-electron-trigger}.
We consider a  $p_T$/$\eta$-independent relative systematic uncertainty of 1\%. 
We generate a $p_T$-dependent systematic uncertainty  by modifying the position of the effective trigger threshold in $p_T$ by 1~GeV.

The resolution and $p_T$ scale for high $p_T$, isolated electron candidates and the associated systematic uncertainties  have been
measured by the ATLAS~\cite{bib:atlas-electronE} and CMS~\cite{bib:cms-electronID} collaborations.
We simulate the resolution in electron $p_T$ by means of a Gaussian with a width $\sigma_{p_T}$ given by
$\frac{\sigma_{p_T}}{p_T} = \frac{0.16}{\sqrt{p_T}}$.
We consider a  $p_T$/$\eta$-independent systematic uncertainty on the resolution by varying the constant term in the formula for
$\frac{\sigma_{p_T}}{p_T}$ by an amount $\pm 0.002$~\cite{bib:atlas-electronEres}.
The $p_T$ scale is calibrated with a precision of around 0.2\% in both ATLAS and CMS~\cite{bib:atlas-electronE,bib:cms-electronID}.
We consider a  $p_T$/$\eta$-independent relative systematic uncertainty of 0.2\% on the $p_T$ scale.
We generate $\eta$-dependent systematic uncertainties  by modifying the resolution and $p_T$ scale only in the endcap region,
whilst the values in the barrel region remain unchanged.

Since we apply tight isolation requirements on the candidate leptons, we make the conservative assumption that the identification 
efficiency for non-prompt electrons from heavy flavour decays is the same as that for prompt electrons.
In~\cite{bib:atlas-electronID2012} around 2/3 of the
background electrons arise from heavy flavour decays.
The remaining background arises from photon conversions and misidentified hadrons, which are difficult to simulate in the context of
our parameterised detector simulation.
We therefore make the approximation that the total background  to the sample of high $p_T$, isolated electrons in the simulated
events is obtained by multiplying the number of electrons from heavy flavour decays by a factor
of 1.5.

\subsection{Simulation of \tauhad\ candidates}
\label{sec:MCtaus}
Of central importance to any measurement of \RW\ at a hadron collider will be an understanding of the efficiencies and backgrounds
associated with the identification of $\tau$ candidates.
A recent paper by the CMS Collaboration~\cite{bib:cms-tau} describes the methods used to identify \tauhad\ candidates at $\sqrt{s}$~=~13 TeV and
details the methods used to determine identification efficiencies, background rates and energy scale, and their associated
systematic uncertainties, using a data set corresponding to \intL~=~36~\invfb.
We choose to simulate the \tauhad\ identification efficiency and fake probabilities corresponding to the ``very-very-tight'' operating point of~\cite{bib:cms-tau}.
The efficiency is around 30\% and is independent of $p_T$.
The misidentification rate
for hadronic jets is around \mbox{$2\times 10^{-3}$} for $p_T\approx 25$~GeV decreasing to around $10^{-4}$ for $p_T\approx 100$~GeV.
The misidentification rate is approximately independent of \modeta~\cite{bib:cms-taueff}.
We simulate also the discriminants against electrons and muons described in~\cite{bib:cms-tau}:
the ``Tight'' discriminant against electrons has a  \tauhad\ identification efficiency of 75\% and a fake probability of $10^{-3}$;
the ``Tight'' discriminant against muons has a  \tauhad\ identification efficiency of 99\% and a fake probability of $1.4\times 10^{-3}$~\cite{bib:cms-tauemu}.

In our study it is particularly important to assign realistic systematic uncertainties on the identification efficiency for  \tauhad\ candidates.
A tag and probe analysis of \Zltau\ events in~\cite{bib:cms-tau} results in a relative uncertainty on the identification efficiency of  \tauhad\ candidates of 5\% for \tauhad\ $p_T$ up to 60~GeV.
Samples of \tt\ events are used in~\cite{bib:cms-tau} to cross check the  efficiency for \tauhad\ $p_T$ up to 100~GeV; a relative uncertainty
of 7\% for $60<p_T<100$~GeV is assigned.
Implicitly this latter analysis assumes that $\mathcal{B}(\Wtaunu)$ takes its SM value, because it relies on a comparison between
the absolute numbers of \ttWWltau\ candidate events
observed in the CMS data and predicted by the MC.
Clearly, for our proposed test of LU using \tt\ events it would not be legitimate to set  data-MC scale factors for \tauhad\ using
\ttWWltau\ events in this way;
this means that it will be difficult to control the $p_T$ dependence of the identification efficiency for  \tauhad\ candidates beyond the
range in \tauhad\ $p_T$ covered by the \Zltau\ event sample. 
We consider a  $p_T$/$\eta$-independent relative systematic uncertainty of 5\%. 
We generate a  $p_T$-dependent systematic uncertainty on the identification efficiency of  \tauhad\ candidates by modifying the efficiency by a relative amount
that varies linearly between 5\% at $p_T = 30$~GeV and 0\% at $p_T = 130$~GeV and above. 
The $p_T$ resolution for \tauhad\ candidates is given by $\frac{\sigma_{p_T}}{p_{T}} = 0.16$~\cite{bib:cms-tau-2012}.

The relative uncertainties on the probabilities for electrons, muons, and hadronic jets to be misidentified as a \tauhad\ candidate are around 10\%~\cite{bib:cms-taumisid}.
The uncertainty on the  $p_T$ scale for \tauhad\ candidates is around 1\%~\cite{bib:cms-tau}.
We consider an $\eta$-dependent systematic uncertainty  by modifying the $p_T$ scale by 1\% only in the endcap region,
whilst the scale in the barrel region remains unchanged.
Preliminary systematic uncertainties of a similar magnitude have been assessed by the ATLAS Collaboration for the identification of
\tauhad\ candidates at  $\sqrt{s}$~=~13~TeV~\cite{bib:atlas-tau2015data}, using the methods described in~\cite{bib:atlas-tau}. 

Backgrounds from MJ events
are typically estimated at hadron colliders using data driven methods.
Such backgrounds cannot reliably be estimated from MC.
Of the four signal samples,  \Zltau\ will be the sample with the largest fraction of MJ background.
An important motivation for our  choice to simulate the ``very-very-tight'' operating point of~\cite{bib:cms-tau} for \tauhad\ identification,
which has relatively low efficiency but high rejection power, is to
minimise the uncertainties arising from MJ backgrounds.
From~\cite{bib:cms-muonID} and~\cite{bib:cms-tau}  we estimate the fraction of MJ background in the selected \Zltau\ sample to be around 5\%, with a
relative uncertainty of around 5\%~\cite{bib:cms-taumj}.

The probability to mis-measure the sign of the charge of lepton candidates is expected to be less than 1\% over the range of $p_T$
of relevance to our study and is neglected in our simulation.

\subsection{Simulation of hadronic jets, including $b$-tagging}
\label{sec:MCjets}
Jet finding is performed at particle-level using the anti-$k_t$ algorithm~\cite{bib:antikt} with a distance parameter $R=0.4$.
Determinations of the jet energy scale (JES) and jet energy resolution (JER), along with their systematic uncertainties,  have been
presented by the ATLAS~\cite{bib:atlas-jes} and CMS~\cite{bib:cms-jes} collaborations.
We simulate the resolution in jet $p_T$ by means of a Gaussian with a width $\sigma_{p_T}$ given by
$\frac{\sigma_{p_T}}{p_T} = \frac{1.0}{\sqrt{p_T}}$.
We assume an uncertainty on the jet energy scale of 1\%.

The efficiencies, backgrounds and systematic uncertainties associated with the $b$-tagging of hadronic jets have been
presented by the ATLAS~\cite{bib:atlas-b-tag} and CMS~\cite{bib:cms-b-tag} collaborations.
We choose to simulate tag probabilities according to those given for the ``DeepCSV Loose'' category of~\cite{bib:cms-b-tag}.
Jets with \modeta~$<$~2.5 are flagged as $b$-tagged with probabilities that depend on their flavour at truth level as follows: truth $b$ 85\%, truth $c$ 40\%, truth
light quark or gluon 1\%.
These tag probabilities are approximately independent of $p_T$ and \modeta.
The relative uncertainty in the $b$-jet efficiency scale factors is around 1.5\% and the
uncertainties in the $c$-jet and light-jet mis-tag scale factors are around 4\% and 10\% respectively~\cite{bib:cms-b-tag-eff}.

The value of $\slashed{E}_T$  is calculated from the vector sum at particle
level of the $p_T$ of all neutrinos in the event.
Resolution in  $\slashed{E}_T$ is taken into account by adding the difference between particle-level and detector level  $p_T$ of
each lepton and jet in the event.
We have checked that this procedure reproduces approximately the $\slashed{E}_T$ resolutions given
in~\cite{bib:atlas-met} and~\cite{bib:cms-met}.

The effects of multiple proton-proton collisions or ``pile-up'' are not simulated in our study.
In general, the lepton and jet identification algorithms employed by ATLAS and CMS are designed to have small pile-up
dependence~\cite{bib:pile-up}.
We take the
performance values we have implemented to represent averages over the pile-up conditions experienced at the LHC.


\begin{thebibliography}{00}
\bibitem{bib:coordinates}
The pseudorapidity, $\eta$, is defined as $\eta = -\ln[\tan(\frac{\theta}{2})]$, where $\theta$ is 
the polar angle with respect to the beam direction, in the laboratory frame.
\bibitem{bib:cms-trigger}
  The efficiency of the single muon trigger has been measured in the barrel (endcap) region of CMS to be around 96\%
(90\%)~\protect\cite{bib:cms-muonID} and to be almost independent of $p_T$.
In our simulation of the single muon trigger efficiency we choose to use the smaller values given in the text for ATLAS.
\bibitem{bib:atlas-elec-eta-crack}
{For example, the region $1.37<|\eta|<1.52$ is usually excluded for precision measurements involving electrons or \tauhad. We simulate this gap in this detector parameterisation.}
\bibitem{bib:atlas-electronIDeff}
The efficiency for electron identification according for the ``Tight likelihood'' category is shown as
functions of $p_T$ and $\eta$ in Figure~8 of~\protect\cite{bib:atlas-electronID}.
\bibitem{bib:atlas-electron-trigger}
We simulate approximately the efficiency for the single-electron trigger according to that given as a
function of $p_T$ in Figure~26~(a) of~\protect\cite{bib:atlas-trigger}, except that the value of $p_T$ on the abscissa is shifted up by
2~GeV to account for the fact that
for most of the LHC run~2 the trigger threshold was set at $p_T = 26$~GeV~\protect\cite{bib:atlas-triggerthreshold-electron}.
\bibitem{bib:cms-taueff}
The \tauhad\ identification efficiencies and jet fake probabilities corresponding to various working points of the MVA-based \tauhad\ isolation algorithm are shown as a function of $p_T$ in Figure~4 of~\protect\cite{bib:cms-tau}.
\bibitem{bib:cms-tauemu}
The \tauhad\ identification efficiencies and electron fake probabilities corresponding to various working points of the MVA-based electron discrimination algorithm are shown as a function of $p_T$ in Figure~5 of~\protect\cite{bib:cms-tau}.
The \tauhad\ identification efficiencies and muon fake probabilities are given in section~5.4 of~\protect\cite{bib:cms-tau}.
\bibitem{bib:cms-taumisid}
The measurement in the CMS data of the probabilities for hadronic jets, electrons, and muons to be misidentified as \tauhad\ candidates and comparison
with MC simulations is described in section~10 of~\protect\cite{bib:cms-tau}.
\bibitem{bib:cms-taumj}
In Figure~7~(left) of~\protect\cite{bib:cms-tau} the fraction of MJ background in the signal region of visible mass in the selected
\Zltau\ sample is seen to be around 20\%, when using the ``tight'' operating point of the MVA-based \tauhad\ identification algorithm and applying the ``tight'' isolation
requirement to the muon.
It can be seen from Figure~4~(left) of~\protect\cite{bib:cms-tau} that the signal efficiency for \tauhad\ decreases by a factor of around 1.5
between the ``tight'' and ``very-very-tight'' operating points and from Figure~4~(right) of~\protect\cite{bib:cms-tau} that the
background rejection improves by a factor of three.
Therefore background to signal ratio improves by roughly a factor of two
between the ``tight'' and ``very-very-tight'' \tauhad\ operating points.
Similarly, in section~6.2 of~\protect\cite{bib:cms-muonID} it is stated that the probability for a muon produced in a hadronic jet
to satisfy tight isolation requirements is about 0.05 in the barrel, and goes up to about 0.15 in the endcap.
This probability is around a factor of two larger than the probability of 0.03 that we have assumed for the probability for a muon produced in a hadronic jet.
Taking these two factors of two into account we can, therefore, estimate the fraction of MJ background in the \Zltau\ sample selected using the ``very-very-tight''
operating point of the MVA-based \tauhad\ identification algorithm and our chosen electron/muon isolation to be around $20/(2\times 2) = 5$\%.
The relative uncertainty on the MJ background quoted in section~9.1 of~\protect\cite{bib:cms-tau} is 5\% and results from the
limited size of the control samples in the \intL~=~36~\invfb\ of CMS data used in the measurement.
\bibitem{bib:cms-b-tag-eff}Tag probabilities  for the ``DeepCSV'' tagging algorithms are shown as
  functions of $p_T$ and $\eta$ in Figure~17 of~\protect\cite{bib:cms-b-tag}.
  The data-simulation scale factors and their uncertainties are shown in Figure~53.
\bibitem{bib:pile-up}
In particular, the efficiency of the DeepCSV Loose b-tagging algorithm of CMS has only a very moderate dependence on the number of
pile-up interactions, as is shown in the top right-hand plot of Figure~17 of~\protect\cite{bib:cms-b-tag}.
This means that the b-tagging requirements in the event selections will not introduce any significant difference in the pile-up
distributions in the selected  \ttWW\ and \Ztautau\  event samples.
\bibitem{bib:ww-tt-z-ratios}
Table 5 of~\protect\cite{bib:atlas-ww} gives a number of predictions of the $W^+W^-$ fiducial cross-section, with systematic uncertainties in the range 2--5\%.
Given that $Z$ boson and EW diboson production are both dominated by \qq-initiated processes it is reasonable to assume that
uncertainties due, e.g., to PDFs would at least partially cancel in the ratio.
We assume an uncertainty of 3\% on the ratio of the cross sections for $Z$ boson and EW diboson production.
In~\protect\cite{bib:cms-tt} the \tt\ cross section is measured with a precision of around 3\%.
Since \tt\ production is dominated by gluon-initiated processes we do not assume any cancellation of systematic uncertainties in
evaluating the ratio to the $Z$ boson cross section.
\bibitem {bib:atlas-zptsyst}	
  The assigned systematic uncertainties in our study correspond to the approximate size of the uncertainties on individual bins in
  \Zpt\ in~\protect\cite{bib:atlas-zptphistar}.
  These are dominated by statistical uncertainties in the relevant region of \Zpt.
\bibitem{bib:binomial}
To calculate the uncertainty from the alternative \tt\ generator-level sample, equation~\protect\ref{eqn:RbbWW} is modified to use
the ratio of the efficiencies for selecting \ttWWemu\ and \ttWWltau\ events.
The advantage of this modification is that the efficiencies have binomial uncertainties, which minimise the effects of statistical fluctuations in the alternative sample.
\end{thebibliography}
\end{document}